\documentclass[10pt,conference]{IEEEtran}
\usepackage[utf8]{inputenc}
\usepackage{times}
\usepackage{verbatim}
\usepackage[noadjust]{cite}
\usepackage{graphicx}
\usepackage{setspace}
\usepackage{color}
\usepackage{authblk}
\usepackage{commath}
\usepackage{tabulary}
\usepackage{subfig}
\usepackage{listings}
\usepackage{amsmath}
\usepackage{tcolorbox}
\usepackage{algorithm}
\usepackage{algorithmic}
\usepackage[figurename=Figure,labelsep=period,tablename=Table]{caption}
\usepackage{xr}
\usepackage{subfiles}

\usepackage{textcomp}
\usepackage[normalem]{ulem}
\usepackage{enumitem}
\setlist{leftmargin=3mm}

{

\author {
	{{Pankaj Saha}, {Angel Beltre}, and {Madhusudhan Govindaraju}}
	\vspace{1.6mm}\\
	\fontsize{10}{10}\selectfont\itshape
    {Cloud and Big Data Laboratory},
	{State University of New York (SUNY) at Binghamton}\\
	\fontsize{9}{9}\selectfont\ttfamily\upshape
	{
		{{$\lbrace$psaha4, abeltre1, mgovinda$\rbrace$@binghamton.edu }}
	}
}}

\begin{document}

\def\sharedaffiliation{%
\end{tabular}
\begin{tabular}{c}}

\title{Tromino: Demand and DRF Aware Multi-Tenant Queue Manager for Apache Mesos Cluster}

\renewcommand{\thetable}{\arabic{table}}
\date{}
\maketitle
\thispagestyle{empty}
\pagestyle{empty}

\begin{abstract}
Apache Mesos, a two-level resource scheduler, provides resource sharing across multiple users in a multi-tenant cluster environment. Computational resources (i.e., CPU, memory, disk, etc. ) are distributed according to the Dominant Resource Fairness (DRF) policy. Mesos frameworks (users) receive resources based on their current usage and are responsible for scheduling their tasks within the allocation. We have observed that multiple frameworks can cause fairness imbalance in a multi-user environment. For example, a greedy framework consuming more than its fair share of resources can deny resource fairness to others.  The user with the least Dominant Share is considered first by the DRF module to get its resource allocation. However, the default DRF implementation, in Apache Mesos' Master allocation module, does not consider the overall resource demands of the tasks in the queue for each user/framework. This lack of  awareness can result in users without any pending task receiving more resource offers while users with a queue of pending tasks starve due to their high dominant shares. 

In a multi-tenant environment, the characteristics of frameworks and workloads must be understood by cluster managers to be able to define fairness based on not only resource share but also resource demand and queue wait time.  We have developed a policy driven queue manager, {\it Tromino}, for an Apache Mesos cluster where tasks for individual frameworks can be scheduled based on each framework's overall resource demands and current resource consumption. Dominant Share and demand awareness of Tromino and scheduling based on these attributes can reduce (1) the impact of unfairness due to a framework specific configuration, and (2) unfair waiting time due to higher resource demand in a pending task queue. In the best case, Tromino can significantly reduce the average waiting time of a framework by using the proposed Demand-DRF aware policy. 
\end{abstract}
\section{introduction}




In clouds and large clusters, several different types of applications are executed and multiple users/groups can demand difference resources to execute their tasks. In such shared environments, {\it Fairness} needs to be defined and maintained. Apache Mesos \cite{the_mesos_paper} is a data center Operating System that combines resources from all participating cluster nodes and provides a global view as a single giant pool of resources. Fairness for multiple resources in this multi-tenant environment is defined using the Dominant Resource Fairness (DRF) policy, introduced by Ghodsi et al.~\cite{the_drf_paper}.

Apache Mesos acts as a resource manager and different Mesos frameworks act as resource consumers. One of the widely known frameworks, Apache Aurora~\cite{apache_aurora_web}, was developed by Twitter for running services and short-lived jobs. Mesosphere developed, Marathon~\cite{marathon_web}, a framework for long-running services and container orchestration. The Chronos~\cite{Cchronos_web} framework was developed for periodic execution of cron jobs. In our previous work, we developed Scylla~\cite{the_scylla_paper}, which is a Mesos framework for running MPI jobs on cloud-based HPC systems. 
Apache Mesos has proven scalability of running on more than 10K nodes\cite{mesos_news_web} in a production setup, and it seamlessly supports Docker~\cite{the_docker_paper} as its primary choice for containerized applications.
   
The introduction of Apache Mesos and its DRF based allocation module led to widespread acceptance by the cloud computing community, as workload fairness and optimal resource utilization are essential for multi-framework execution environments. In our previous work~\cite{saha_drf_explore_paper}, we identified how resource allocation and fairness could be affected due to framework settings such as offer refusal period, resource holding period, task arrival rate, and second level scheduling.  Each framework in a Mesos cluster is typically designed for a specific type of application, but its configuration properties can hinder fairness and induce starvation in a cluster. 

To observe the unfairness in an Apache Mesos cluster, we set up a cluster environment of 4 nodes where each node contains \(< 8\ CPU, 16\ GB\ memory >\) of resources. We orchestrated synthetic jobs, launched by Scylla and Marathon, wherein each required \( < 1\ CPU , 2\ GB\  memory>\) of resources. 
In an ideal fair distribution scenario, each framework should be able to run 16 jobs each. As each job is identical in terms of resource requirements, the number of jobs launched by each framework is proportional to the amount of resources consumed by each framework.

In Figure \ref{fig_moti_scylla_marathon_dotted}, we can observe how Marathon utilizes more resources and launches more tasks in the cluster while Scylla uses comparatively low amount of resources. 
We measure the unfairness \text{$U_{A}$} to framework A by using the following formula proposed in earlier work~\cite{saha_drf_explore_paper}:
\[ U_{A} = (\dfrac{Area_{i,j} \ by \ framework_{A}}{{Area_{i,j} \ by \ fair \ graph}})*100\]

\( Area_{i,j} \) is the area under the curve from point i to j

In Figure \ref{fig_moti_scylla_marathon_dotted}, the dotted horizontal line shows the fairness baseline, which indicates the number of tasks each framework should be able to execute in a fair distribution manner. Two vertical dotted lines represent the beginning and end of the period for which we have calculated the fairness.


In Figure \ref{fig_allocation_cycle}, the flow diagram shows how the Mesos allocation module distributes resources to multiple frameworks in a Mesos cluster. Apache Mesos' implementation of DRF does not consider the overall resource demands of all the tasks pending in each framework's queue. While allocating resources, it only considers the current resource consumption of each framework. This can lead to a situation where a framework with a higher number of tasks in its queue faces an extended waiting time for the tasks to be launched. This phenomenon can increase the cluster's overall waiting time by imposing unfair waiting time to frameworks with higher demands.While allocating resources, the Mesos Master picks agent nodes with available resources in a random order. It does not validate if the available resources are useful to a framework, or if they are aligned with the resource demands. We have used of-the-shelf allocation module of Apache Mesos to study this phenomenon and present schemes on how demand awareness can be achieved while allocating resources.

 We have developed a queue manager, \textit{Tromino}, which is aware of DRF and the dominant share of each framework in the cluster. Tromino controls the waiting task queue of each framework and releases tasks based on the dominant share for better fairness. We have also considered a situation where few frameworks in a cluster have higher demands compared to others. Releasing tasks only based on the dominant share may improve resource fairness but could also increase the total waiting time of tasks for that framework.

The key contributions of this work are the following:
\begin{itemize}
\item We have designed and developed a queue manager, {\it Tromino}, on top of the Apache Mesos scheduler, which keeps track of incoming tasks of all the registered frameworks in the cluster and their current resource consumption.
\item {\it Tromino} monitors the resource demands and resource consumption information from the waiting queue of jobs and Mesos Master respectively to control task dispatching.
\item We have shown how our demand aware scheduling on top of Mesos’ default DRF, can reduce the average waiting time across all frameworks.
\item We present and show how different policies of task dispatching, based on the demand and dominant share, affect fairness in four different case scenarios.
\end{itemize}

\begin{figure}[h!] 
\vspace{-0.1em}
  \includegraphics[width=0.5\textwidth]
  {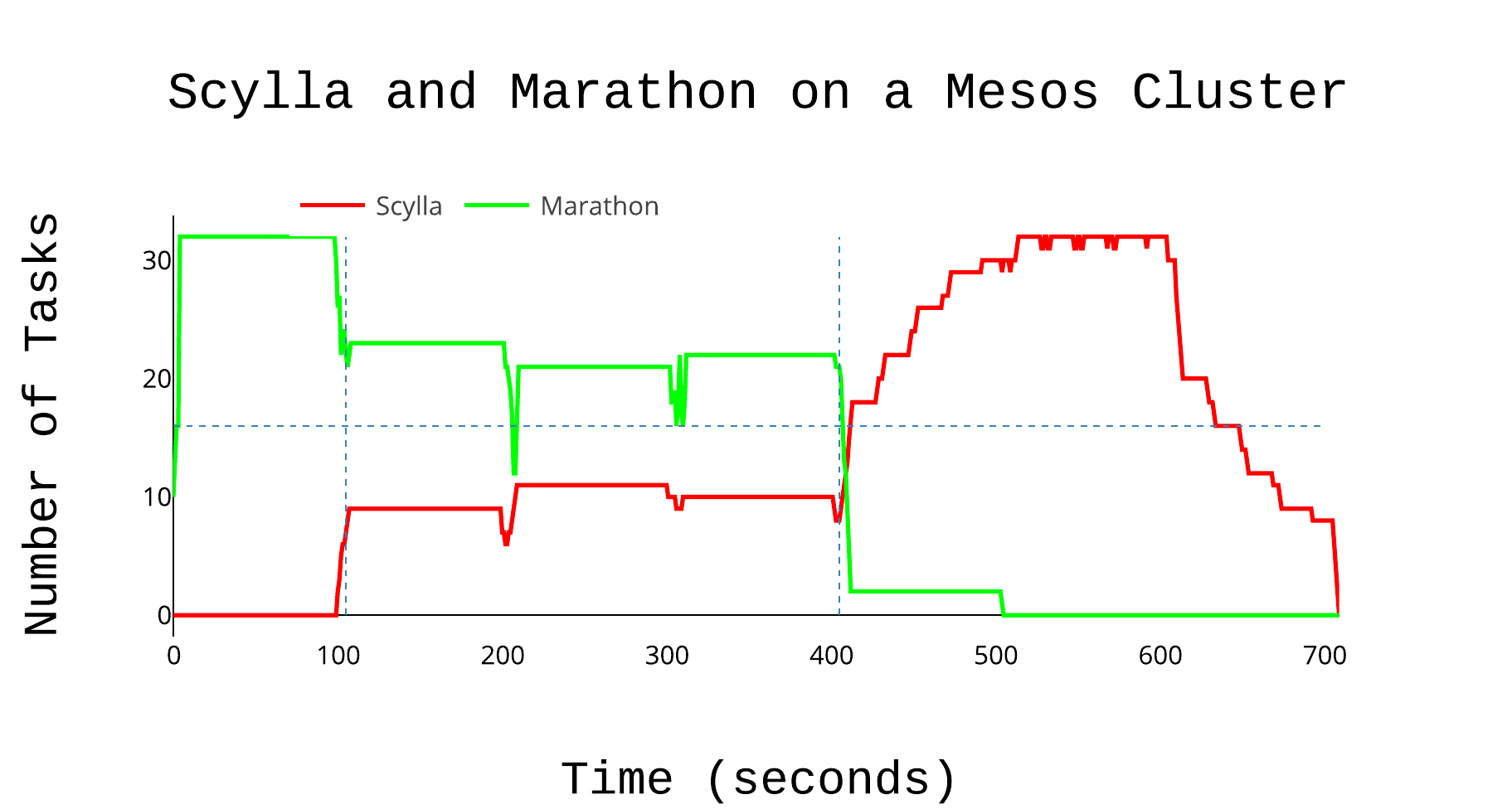}
\caption{{\it Scylla and Marathon are chasing for resources in a Mesos cluster. Marathon is able to launch several more tasks than Scylla. Scylla's tasks face longer wait times due to unfair distribution.}}
\label{fig_moti_scylla_marathon_dotted}
\vspace{-0.7em}
\end{figure}

\section{background}
\subsection{Apache Mesos}\label{apache_mesos}

Figure \ref{fig_mesos_architecture} shows the architectural components of Apache Mesos. It consists of three major components. Mesos Agent, Mesos Framework, and Mesos Master. \textbf{Mesos Agent} consists of the computational resources like CPU, memory, disk, etc. that are required to execute tasks. Each Mesos Agent needs to have a Mesos Executor installed to receive task execution requests. \textbf{Mesos executor} is a program that resides in all the Mesos agent nodes and executes tasks upon requests from the Mesos master. \textbf{Mesos Frameworks} are the users that have a pending queue of tasks to be launched, along with user-defined resource demands. They also have a scheduler that decides the task that has to be launched on an each agent node after resources are offered to them. During framework registration, each framework needs to provide the executable path of the Mesos Executor, or else the default Mesos Executor takes care of the requested tasks. For example, Apache Aurora uses Thermos~\cite{thermos_web} as the Mesos Executor whereas Mesosphere Marathon~\cite{marathon_web}, developed by the core developers of Apache Mesos, uses the default Mesos Executor. \textbf{Mesos Master} negotiates between the Mesos frameworks and Mesos agents to allocate resources based on current resource consumption by each framework. In a distributed production setup, multiple Mesos Masters are installed, and one of them is elected as a leader by zookeeper~\cite{the_zookeeper_paper} to serve as the resource broker for the cluster. Mesos Master consists of a resource allocation module, which decides to allocate resources to each framework periodically based on the DRF policy~\cite{the_drf_paper}. The Mesos setup follows the following steps to allocate resources for executing tasks on the agent nodes.

\begin{itemize}
\item \textit{\textbf{Step 1 - Advertising Resources}}. At the beginning of an allocation cycle each Mesos Agent advertises its available resources like CPU, memory, disk, etc. to the Mesos Master.
\item \textit{\textbf{Step 2 -  DRF based Resource Allocation}}. Based on the current resource consumption by each framework, the Mesos Master's allocation module decides the resource allocation for each framework for executing the tasks. The Mesos Master does not take into account the resource needs of a framework before sending it offers. This step is considered as the 1st level scheduling in a Mesos setup.
\item \textit{\textbf{Steps 3 - Generating Matching List of Tasks and Agents}}. Now, each framework decides how to schedule tasks across the resources in the agent nodes allocated to it. The framework takes into account the hardware, device, or other task specific constraints provided by the user to the framework. 
Once a framework makes its decision on using or rejecting the resources from each allocated agent, it makes a list of tasks matching with the Mesos agents and sends it to the Mesos Master. The matching of resources offered by the Mesos Master to the requirements of tasks in a framework is called the 2nd level scheduling. 
\item \textit{\textbf{Step 4 - Assigning Tasks to Allocated Agents}}. If the framework's request for resources for each task does not exceed the available resources in the agent nodes, the Mesos Master request Mesos agents to execute the task. If the frameworks' resource requirements do not match the availability on the agent nodes, the execution request is not sent to the agents. The resources that are not used by each framework are returned to the pool of available resources and offered during the next allocation cycle as explained in Figure \ref{fig_allocation_cycle}.
\end{itemize}

\begin{figure}[h!]
\vspace{-0.1em}
  \includegraphics[width=0.5\textwidth]
  {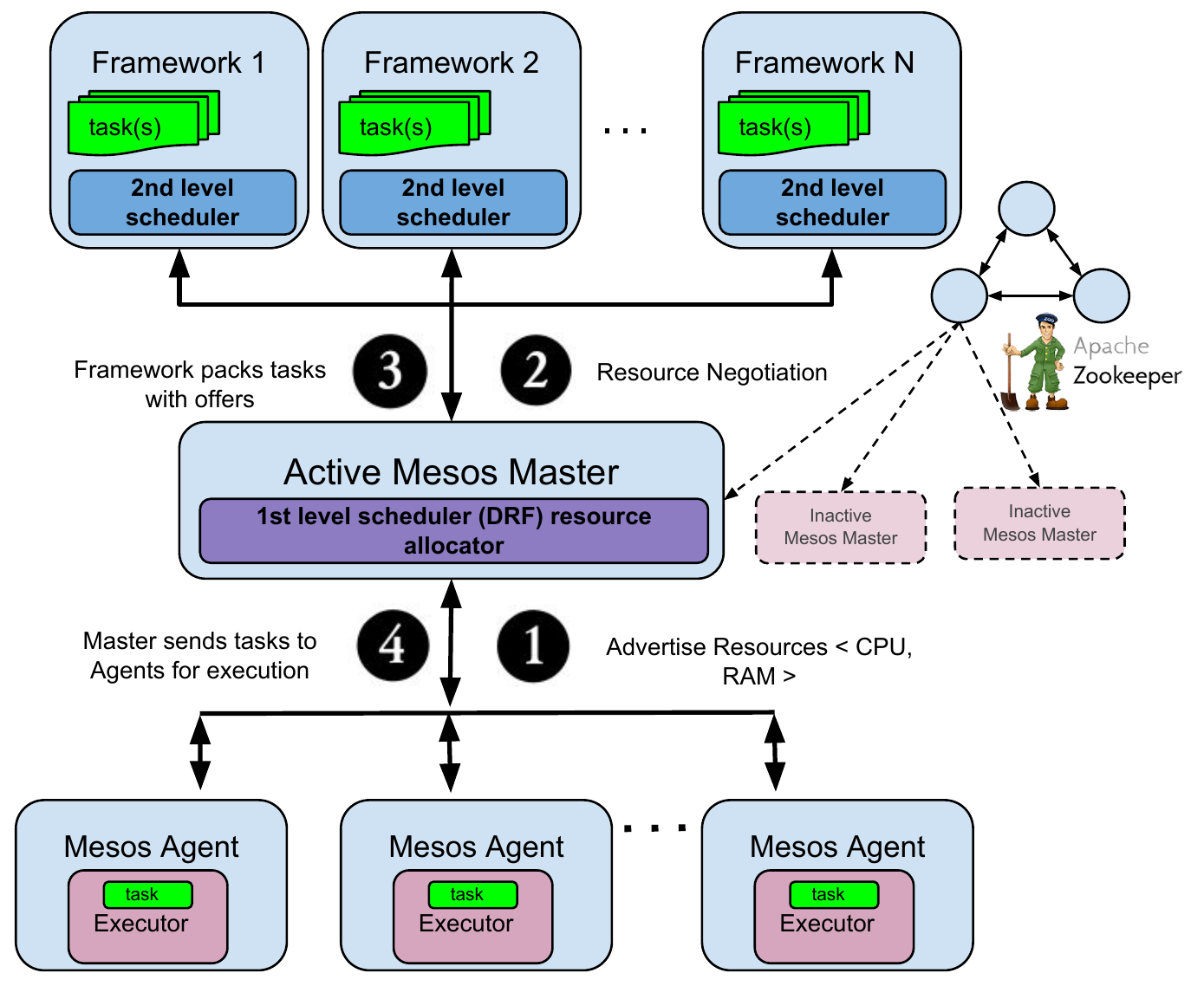}
  \caption{{\it \textbf{Apache Mesos Architecture}: This diagram shows the primary architectural components and resource allocation steps from Mesos Agents to Mesos Framework (user) through DRF based resource allocation by Mesos Master.}}
  \label{fig_mesos_architecture}
   \vspace{-0.7em}
\end{figure}

\subsection{DRF and Apache Mesos} \label{sec:DRF_and_Apache_Mesos}
Resource allocation policies, such as Max-Min, or its more generalized version like the weighted Max-Min, can provide fairness to multiple users in a multi-tenant environment. However, they are designed for a single type of resource. For multiple resources, slot based allocation has became popular with YARN \cite{the_yarn_paper} for Hadoop and map-reduce tasks. However, the slot based allocation has a shortcoming of over or under allocation of resources. In cloud and modern cluster computing environments users can request different types of resources. Multiple jobs can be co-scheduled on the same physical node. The Dominant Resource Fairness (DRF) \cite{the_drf_paper} algorithm was introduced to bring fairness among multiple users competing for various kinds of resources. Apache Mesos is one of the leading cluster resource managers to incorporate DRF. Its resource allocation module is based on DRF.

\begin{figure}[h!]
\vspace{-0.1em}
  \includegraphics[width=0.5\textwidth]
  {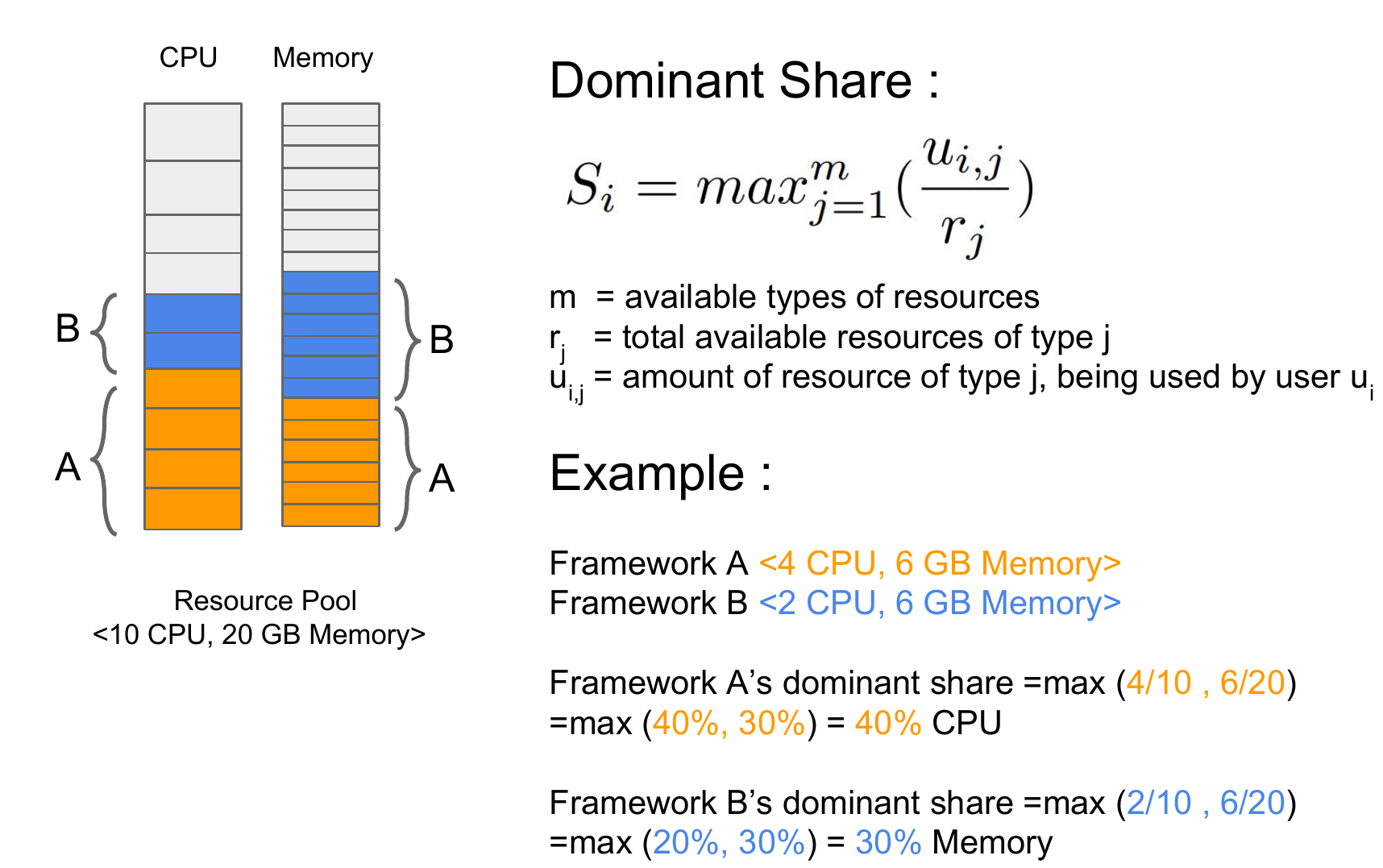}
  \caption{{\it \textbf{Dominant Share}: This diagram pictorially represents the concept of Dominant Share, which decides the available resource allocation to each framework, while consuming multiple types of resources to execute pending tasks in the queue.}}
  \label{fig_dominant_share}
   \vspace{-0.7em}
\end{figure}

To explain how DRF works in Apache Mesos, we illustrate using a simple example how the dominant resource and dominant share are calculated. Figure \ref{fig_dominant_share} shows a pool of computing resources and two frameworks competing for different amount of resources for various tasks in their own queues. Framework A is currently consuming 4 CPUs and 6 GB of memory for all its running tasks. Similarly, Framework B's tasks are consuming 2 CPUs and 6 GB of memory. The total pool of resources in this example consists of 10CPUs and 20 GB memory. Figure~\ref{fig_dominant_share} shows how the dominant share and dominant resource are determined for both the frameworks. The flowchart in Figure~\ref{fig_allocation_cycle} explains how DRF is implemented in Apache Mesos and how the allocation cycle works.

\begin{figure}[h!]
\vspace{-0.1em}
  \includegraphics[width=0.5\textwidth,height=9cm]
  {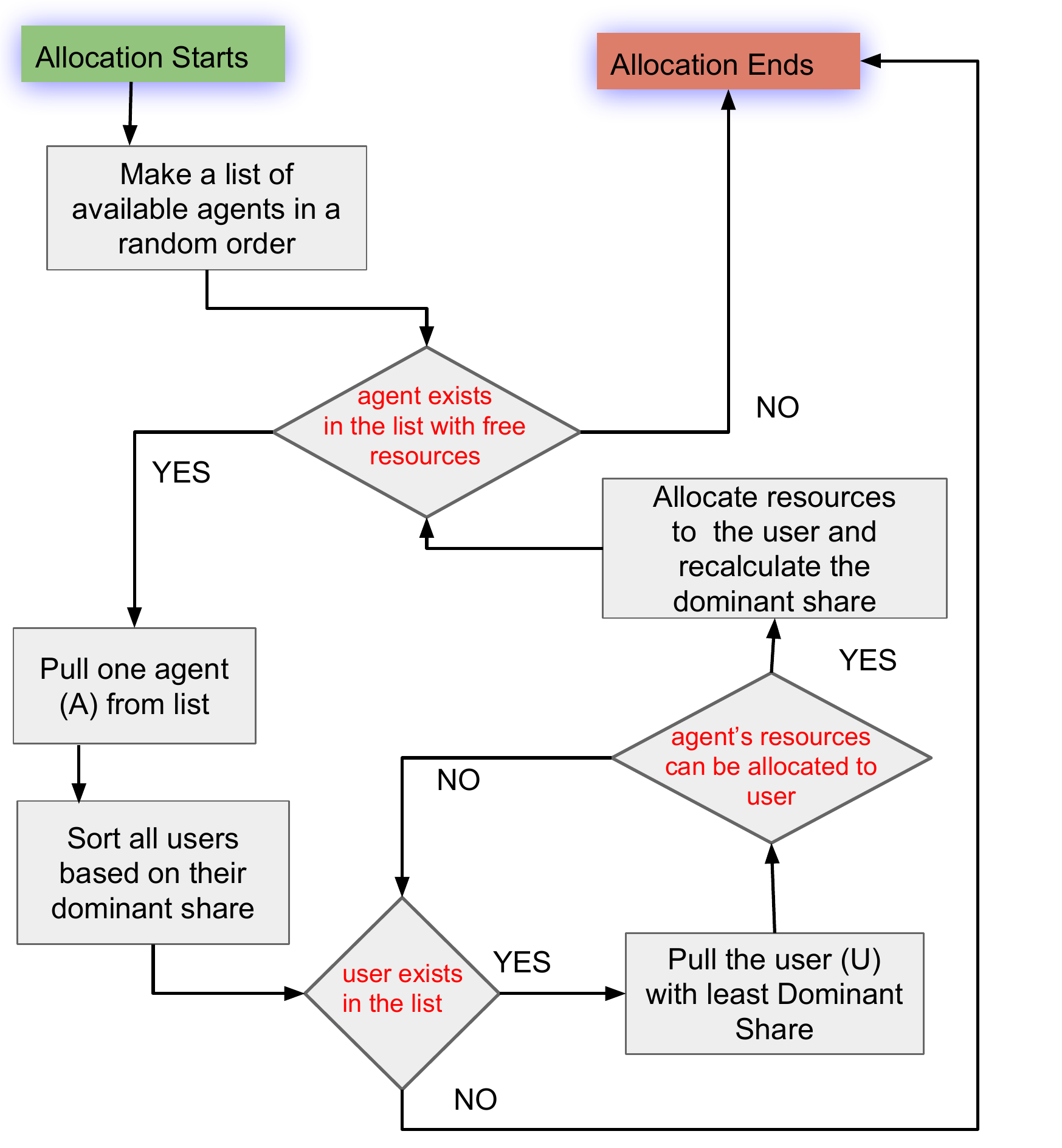}
  \caption{{\it \textbf{Resource Allocation Cycle}: Periodic resource allocation cycle by Mesos Master's allocation module to allocate computational resources from Mesos Agents to Mesos Frameworks determined by dominant share of each framework (user).}}
  \label{fig_allocation_cycle}
   \vspace{-0.7em}
\end{figure}

\section{ Tromino architecture and Strategy }

\subsection {Tromino}
Figure \ref{fig_tromino_architecture} shows how the {\it Tromino} queue manager fits in a Mesos setup to manage the task queues for several Mesos Frameworks. {\it Tromino} fetches cluster and task information periodically from the Mesos Master and keeps track of the tasks in the queue for each framework. In a conventional Mesos setup, the end user submits tasks directly through the frameworks, and each framework's tasks are executed using the steps listed in Section \ref{apache_mesos}. Unlike a conventional setup, in the presence of {\it Tromino}, a user submits tasks directly to {\it Tromino}. {\it Tromino} maintains separate queues for each framework. {\it Tromino} takes into account the following information to decide the task and framework to dispatch: (1) current resource consumption of each framework; (2) the total available resources in the cluster; and (3) the resource demands of tasks in each framework's queue.

\begin{figure}[h!]
\vspace{-0.1em}
  \includegraphics[width=0.5\textwidth]
  {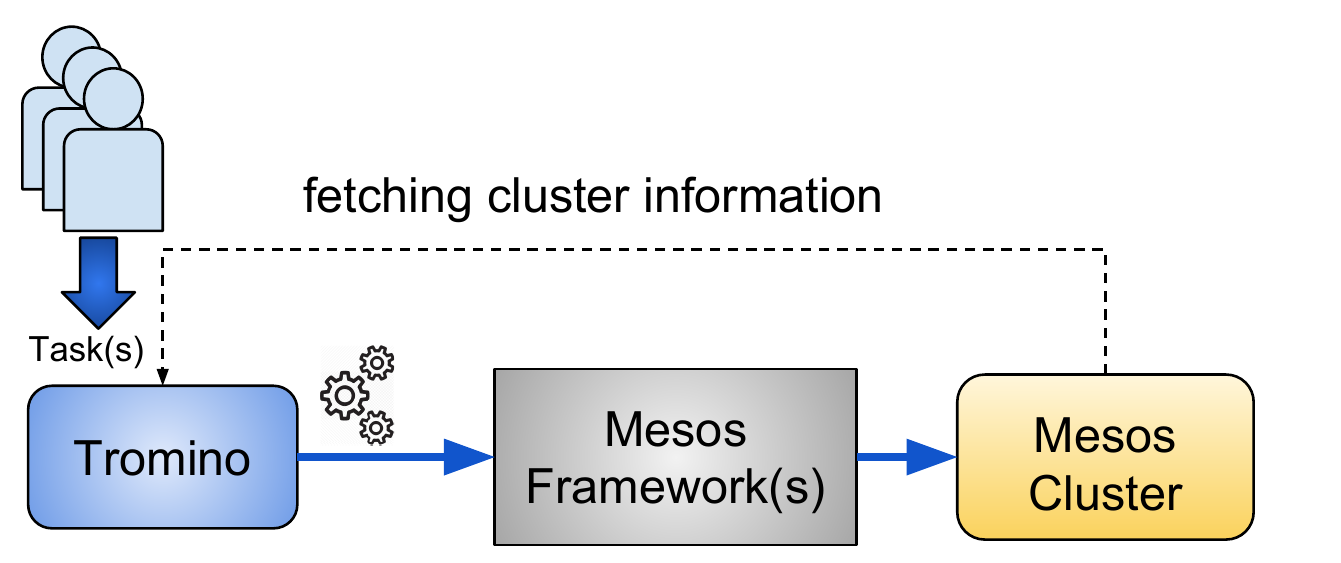}
  \caption{{\it \textbf{Tromino Architecture}: Tromino communicates with the Apache Mesos Master to understand the current resource consumption of each framework and based on the chosen policy it releases tasks from the queue associated with each framework.}}
  \label{fig_tromino_architecture}
   \vspace{-0.7em}
\end{figure}

\subsection{Tromino Manager}
Figure \ref{fig_tromino_manager} shows the components and flow involved in dispatching tasks through {\it Tromino}. {\it Tromino} consists of three major elements (1) Tromino Dispatcher, (2) Tromino Manager and (3) Tromino Scheduler. 

\textit{Tromino Dispatcher} consists of a dispatcher and a task queue for each framework registered with the Mesos cluster. Based on the user's preference for a framework as specified to the Tromino client, {\it Tromino} moves the task to the appropriate dispatcher. Each dispatcher collects information on all the resource demands of the tasks in its queue and the current dominant resource demand of the queue.

\textit{Tromino Manager} periodically communicates with the Mesos Master to fetch information regarding resource consumption of all the frameworks, the dominant share of each framework, and the available resources in the cluster.

\textit{Tromino Scheduler} controls the release of the tasks from each dispatcher's queue to the corresponding frameworks. The tasks are released based on the chosen scheduling policy (see Section \ref{policies}). It consults with the Tromino Manager to decide how many tasks need to be released. 

\begin{figure}[h!]
\vspace{-0.1em}
  \includegraphics[width=0.5\textwidth]
  {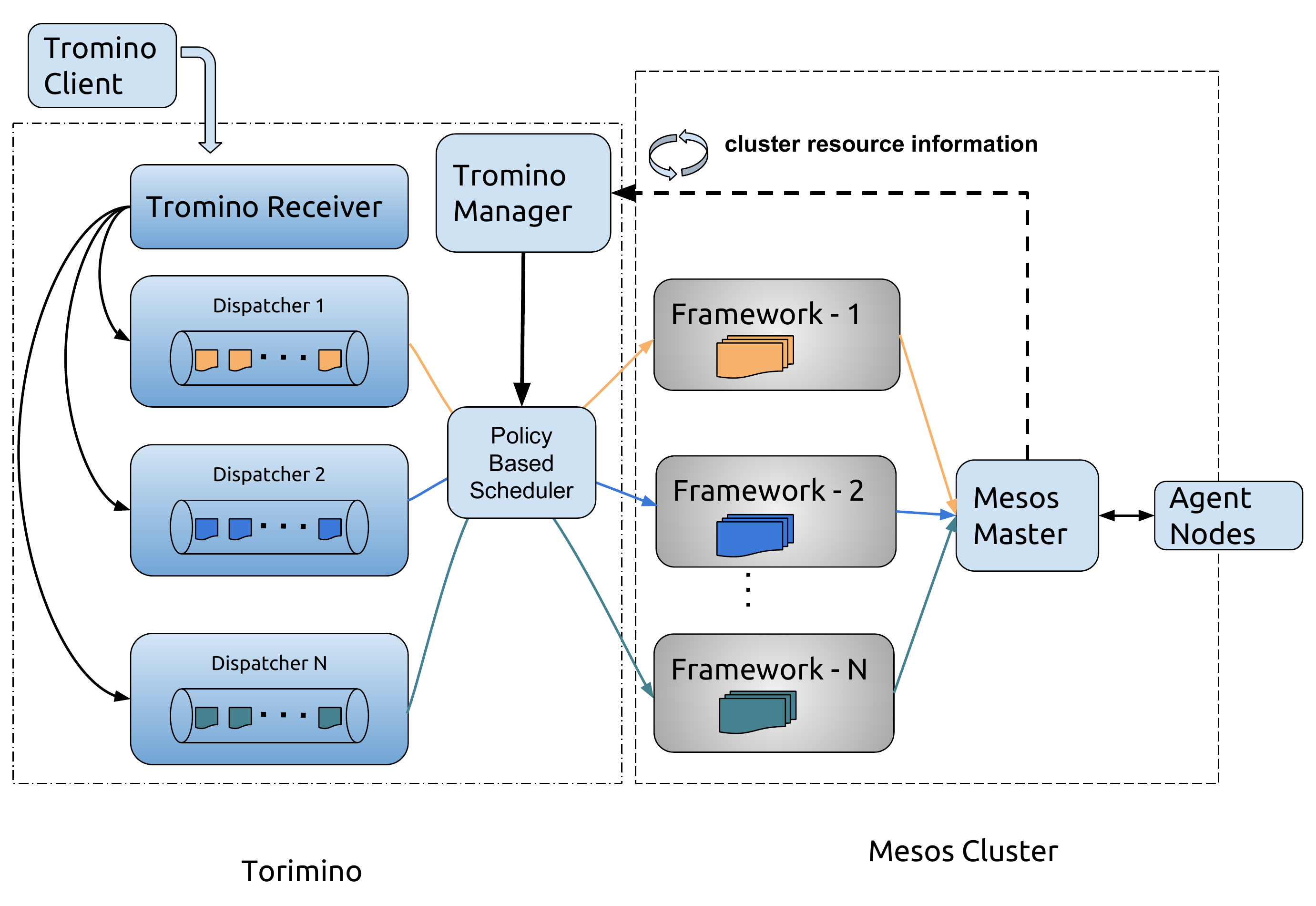}
  \caption{{\it \textbf{Tromino Manager}: Tromino Managers consists of multiple Tromino Dispatchers, one for each framework, and it can communicate with Mesos Master to get information about current resource consumption of each registered active framework. Tromino Manager also communicates with each dispatcher to understand the current resource demand to make decisions regarding the release of tasks from each dispatcher.}}
  \label{fig_tromino_manager}
   \vspace{-0.7em}
\end{figure}

\subsection{Tromino Policies} \label{policies}
We have designed three scheduling policies for the Tromino Scheduler: DRF Aware Policy, Demand Aware Policy, and Demand-DRF Aware Policy. These policies can be extended further based on the scheduling needs of users and applications. In Section \ref{sec:DRF_and_Apache_Mesos}, we discused how the Dominant Share (DS) is calculated for any DRF based algorithm. We introduce the Dominant Demand Share (DDS) attribute in this section. Later in this section, we explain how the Tromino policies use the DS and DDS values.

For example, let us consider a cluster with a total of 20 CPUs and 40 GB of memory, where two frameworks (Framework A and Framework B) are competing for shared resources. Each of the frameworks can have a different number of tasks waiting in their queues to be dispatched. In this example, Framework A  has 10 tasks each with  \( < 1\ CPU, 4\ GB\ memory> \) as the resource demands. Framework B has a total of 5 tasks each with \(< 2\ CPU, 1\ GB\ memory >\) demand waiting in the queue to be dispatched. In Table~\ref{DDS0}, we present how the calculation is carried out for the Dominant Demand Shares (\( DDS_{A}\ and\ DDS_{B}\)) and the dominant demand of each framework. 

\begin{table}[h!]
\centering
\begin{tabular}{ccc}
\begin{tcolorbox}
\vspace{-1em}
\[ DDS_{A} = max [(10*1)/20 , (10*4)/40] = max[0.5,\ \textbf{1.0}] \] 
\[ DDS_{B} = max [(5*2)/20 , (5*1)/40] = max[\textbf{0.5},\ 0.125] \]
\end{tcolorbox}  \\
\end{tabular}
\caption{\textit{Dominant Demand Share (DDS) calculation for both frameworks in the example, before Tromino starts dispatching any tasks to the Mesos cluster.}}
\label{DDS0}
\end{table}
\vspace{-0.5em}

Also, let us consider that Framework A is executing 3 tasks each consuming \( < 1\ CPU, 4\ GB\ memory> \) of resources, and Framework B is executing 5 tasks wherein each is consuming \(< 2\ CPU, 1\ GB\ memory >\) of resources. Now, the Dominant Shares (\( DS_{A}\ and\ DS_{B}\)) for both frameworks and their dominant resources are shown in Table \ref{DS0}.
\begin{table}[h!]
\centering
\begin{tabular}{ccc}
\begin{tcolorbox}
\vspace{-1em}
\[ DS_{A} = max [(3*1)/20 , (3*4)/40] = max[0.15,\ \textbf{0.3}] \] 
\[ DS_{B} = max [(5*2)/20 , (5*1)/40] = max[\textbf{0.5},\ 0.125] \]
\end{tcolorbox} \\
\end{tabular}
\caption{\textit{Dominant Share (DS) calculation for both frameworks in the example, before Tromino starts dispatching any tasks to the Mesos cluster.}}
\label{DS0}
\end{table}
\vspace{-0.5em}

In Table \ref{DDS0} and \ref{DS0}, we show the calculation of DDS and DS for both the frameworks. For Framework A, the values are 1.0 and 0.3 respectively. Similarly, for Framework B, the values are 0.5 and 0.5 for DDS and DS respectively.

\begin{itemize}
\item \textit{\textbf{DRF Aware Policy.}} In this policy, we assign a higher priority to the framework with lesser dominant share and let its corresponding dispatcher release a task. After a task is dispatched, {\it Tromino} recalculates the dominant share and decides the next dispatcher from which a task can be released. For example, as shown in Table \ref{DDS0}, Tromino allows Framework A to release the task. After the first task is dispatched, the DS for Framework B becomes \textbf{0.4}. Subsequently, {\it Tromino} allows another two tasks to be released from Framework A's dispatcher until its dominant share becomes 0.6, which is higher than the dominant share of Framework B.
\begin{table}[h!]
\centering
\begin{tabular}{ccc}
\begin{tcolorbox}
\vspace{-1em}
\[ DS_{A} = max [(6*1)/20 , (6*4)/40] = max[0.3,\ \textbf{0.6}] \] 
\[ DS_{B} = max [(5*2)/20 , (5*1)/40] = max[\textbf{0.5},\ 0.12] \] 
\end{tcolorbox} \\
\end{tabular}
\caption{\textit{Dominant Share of both frameworks after Tromino dispatches 3 tasks from Framework A's dispatcher.}}
\label{DS1}
\end{table}
Now, {\it Tromino} allows two more tasks from Framework B's dispatcher to be released and then the dominant share for both the frameworks is as shown in Table~\ref{DS2}.
\begin{table}[h!]
\centering
\begin{tabular}{ccc}
\begin{tcolorbox}
\vspace{-1em}
\[ DS_{A} = max [(6*1)/20 , (6*4)/40] = max[0.3,\ \textbf{0.6}] \] 
\[ DS_{B} = max [(7*2)/20 , (7*1)/40] = max[\textbf{0.7},\ 0.15] \]
\end{tcolorbox} \\
\end{tabular}
\caption{\textit{Dominant Share after Tromino dispatches 2 more tasks from Framework B's dispatcher until no more resources are available in the cluster.}}
\label{DS2}
\end{table}
At this point, {\it Tromino} stops from any further dispatching as there are no more resources available in the cluster. Finally, {\it Tromino} follows the same steps in the next dispatching cycle if more resources become available.

\item \textit{\textbf{Demand Aware Policy.}}
In this policy, we consider the Dominant Demand Share (DDS) to control the dispatching of tasks from each framework's dispatcher. The framework that has more demand in terms of Dominant Demand Share is given higher priority to dispatch its tasks first. Then, every time a task is dispatched, {\it Tromino} recalculates the DDS and decides which dispatcher gets a chance to release the next task. We observe that in the example discussed in Table \ref{DDS0},  Framework A has higher demand compared to Framework B. In that particular case scenario, {\it Tromino} allows the dispatcher corresponding to Framework A to dispatch the task. It cycles until Framework A dispatches 5 more tasks from its dispatcher queue. At this point, Table \ref{DDS1} shows the DDS for both Frameworks. Now, both the Frameworks have similar DDS, but Framework A cannot launch any tasks as its resource demands cannot be satisfied with the available resources in the cluster. Thus, Framework B's dispatcher dispatches one task from the queue. {\it Tromino} stops this cycle and waits for resources to once again become available so that they could be used in the next cycle. After this cycle, the DDS for both frameworks are presented in Table \ref{DDS2}. In the next dispatching cycle, Framework A may get priority if it still has a higher DDS than Framework B. The DDS of Framework B may go up if it gets new tasks before the next cycle.
\begin{table}[h!]
\centering
\begin{tabular}{ccc}
\begin{tcolorbox}
\vspace{-1em}
\[ DDS_{A} = max [(5*1)/20 , (5*4)/40] = max[0.25,\ \textbf{0.5}] \] 
\[ DDS_{B} = max [(5*2)/20 , (5*1)/40] = max[\textbf{0.5},\ 0.12] \]
\end{tcolorbox} \\
\end{tabular}
\caption{\textit{Dominant Demand Share after Tromino dispatches 5 more tasks from Framework A's dispatcher.}}
\label{DDS1}
\end{table}
\begin{table}[h!]
\centering
\begin{tabular}{ccc}
\begin{tcolorbox}
\vspace{-1em}
\[ DDS_{A} = max [(5*1)/20 , (5*4)/40] = max[0.25,\ \textbf{0.5}] \] 
\[ DDS_{B} = max [(4*2)/20 , (4*1)/40] = max[\textbf{0.4},\ 0.1] \]
\end{tcolorbox} \\
\end{tabular}
\caption{\textit{Dominant Demand Share after Tromino dispatches 5 tasks from Framework A's dispatcher and 1 task from Framework B's dispatcher.}}
\label{DDS2}
\end{table}

\item \textit{\textbf{ Demand and DRF Aware Policy.} }
In this approach, we consider both the demands of each framework and their dominant share. Scheduling based just on the demand may cause unfairness. A framework could end up consuming the entire cluster due to its higher demand while another framework that has significantly fewer number of tasks to execute could starve for resources. We have combined both the dominant share and dominant demand share to generate a Demand-DRF factor in each cycle to decide the number of tasks to be dispatched from each framework's dispatcher.
\end{itemize}

\section{experimental results and evaluation} 
\label{evaluation}

\begin{table}[h!]
\centering
\begin{tabular}{|l|p{0.60\linewidth}|} 
 \hline
 {\bf Software} & {\bf Version}\\
 \hline
 Ubuntu 		& Ubuntu 16.04.2 LTS (Xenial)\\ 
 \hline
 Apache Aurora 	& 0.17.0\\
 \hline
 Marathon       & 1.4.0\\
 \hline
 Apache Mesos 	& 1.3.0\\
 \hline 
\end{tabular}
\smallskip
\label{Software Stack and Vession}
\caption{\textit{Software Stack and Version}}
\label{table_software_stack}
\vspace{-0.5em}
\end{table}

For our experimental setup, we have considered two widely known Mesos frameworks, Apache Aurora and Marathon, along with Scylla, a framework developed by our team. The cluster consists of 8 nodes each with 8 CPUs and 16 GB of memory. We have instrumented {\it Tromino} to receive tasks for Apache Aurora, Marathon, and Scylla at a different task arrival rate. We have kept  the resource requirements of each task identical (i.e., \(<0.5\ CPU, 1\ GB\ memory>\)). The cluster at its peak utilization can execute 128 tasks with such requirements. As all the tasks are identical, the number of tasks that each framework is executing at any instance of time is proportional to the amount of resources consumed by that framework.  Our aim with the experiments is to understand the way resource fairness and task awaiting time varies in different case scenarios. Also, we want to examine how {\it Tromino} policies can achieve better cluster-wide fairness and a reduced average waiting time, over Mesos' default DRF implementation. Our experimental results show the unfairness caused in a Mesos cluster and quantify the fairness in terms of average waiting time for different case scenarios. 

\subsection{\textbf{Experiment 1.} Framework with default configurations and different arrival rates.}
In this experiment, we present a case scenario where Mesos' default DRF based allocation fails to provide cluster-wide fairness due to each framework's varying attributes and task arrival rates. We have instrumented {\it Tromino} to receive tasks for Aurora at a slower rate and receive tasks for Scylla at a higher frequency.  Aurora's default implementation enforces holding resources for a period of time for better scheduling of tasks. On the other hand, Marathon is configured with a relatively greedier second level scheduling policy compared to Aurora. The second level scheduling can significantly affect a framework's individual resource utilization. In our particular case scenario, due to a greedier scheduling policy, Marathon is able to orchestrate more tasks upon receiving resources offers from the Mesos Master. So, Marathon's greedier scheduling policy, and Aurora's characteristic of holding on to resources, affects Aurora because it struggles to launch a fair number of tasks. For Aurora, holding resources makes its Dominant Share stay higher, even though the resources are not used for scheduling tasks. Unlike Aurora, Scylla does not hold resources, and the second level scheduling policy is less greedy compared to Marathon. In Table \ref{table_DRFAwarePolicy}, we show the number of tasks for each framework along with task arrival rate and attributes that can impact the resource distribution.


\begin{table}[H]
\begin{tabular}{|l|l|l|l|l}
\cline{1-4}
\textbf{} & {\bf \# of tasks} & \begin{tabular}[c]{@{}l@{}}{\bf Arrival Rate}\\  (sec)\end{tabular} & {\bf Attribute} &  \\ \cline{1-4}
Marathon & 1000 & 1 & Bin-Packing &  \\ \cline{1-4}
Scylla & 700 & 1.5 & -- &  \\ \cline{1-4}
Aurora & 500 & 2 & Holds resources &  \\ \cline{1-4}
\end{tabular}
\caption {\it Configuration: Aurora, Marathon and Scylla with different task arrival rate along with attributes that may affect the resource fairness in the cluster.} 
\label{table_DRFAwarePolicy}
\vspace{-0.5em}
\end{table}

%
\begin{figure}[h!]
\vspace{-0.5em}
  \includegraphics[width=0.5\textwidth]
  {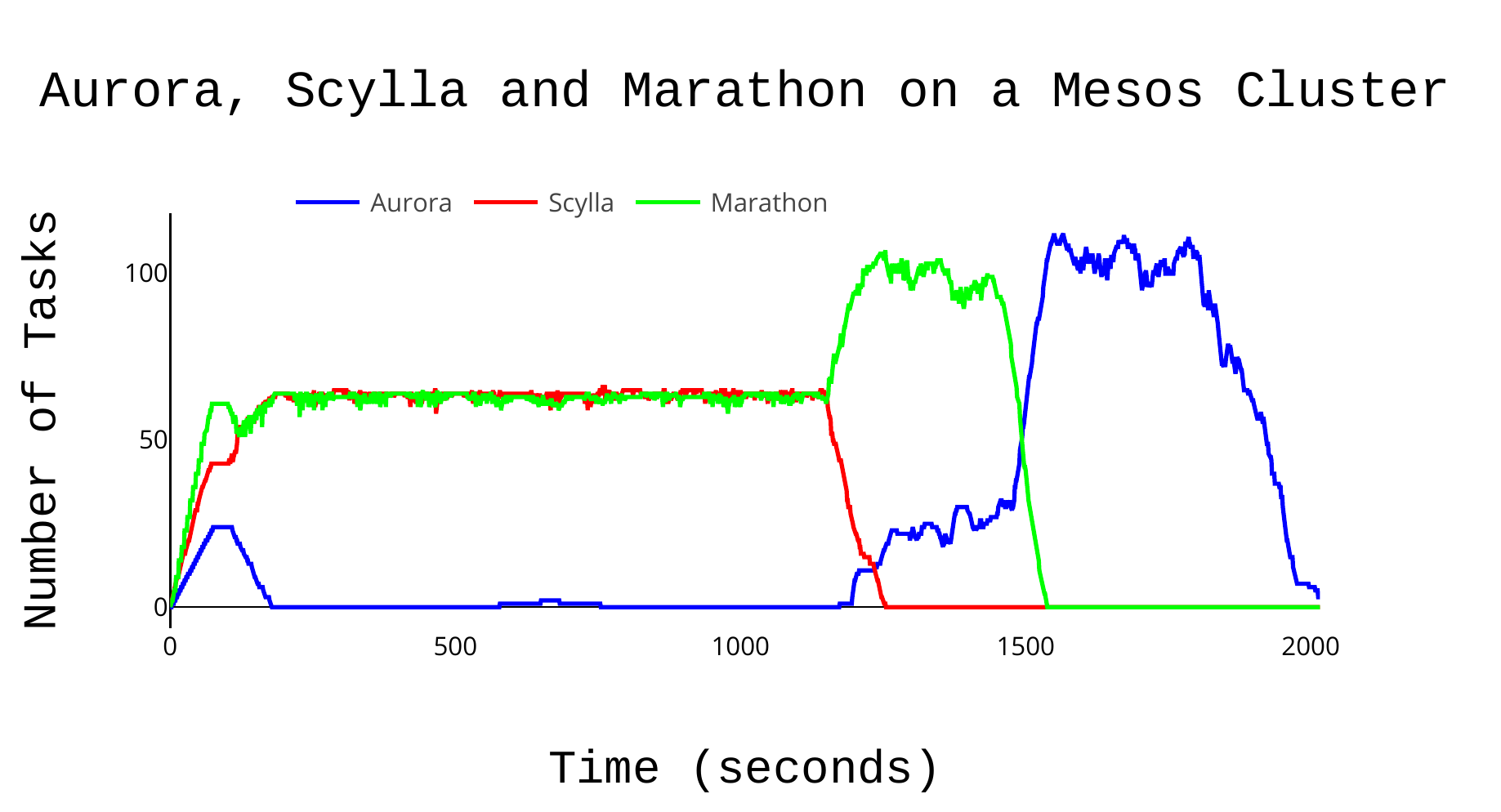}
  \caption{{\it Aurora is not able to launch its pending tasks until Marathon and Scylla are done with executing their tasks.}}
  \label{fig_msa_nosense}
   \vspace{-0.7em}
\end{figure}

\begin{figure}[h!]
\vspace{-0.1em}
  \includegraphics[width=0.5\textwidth]
  {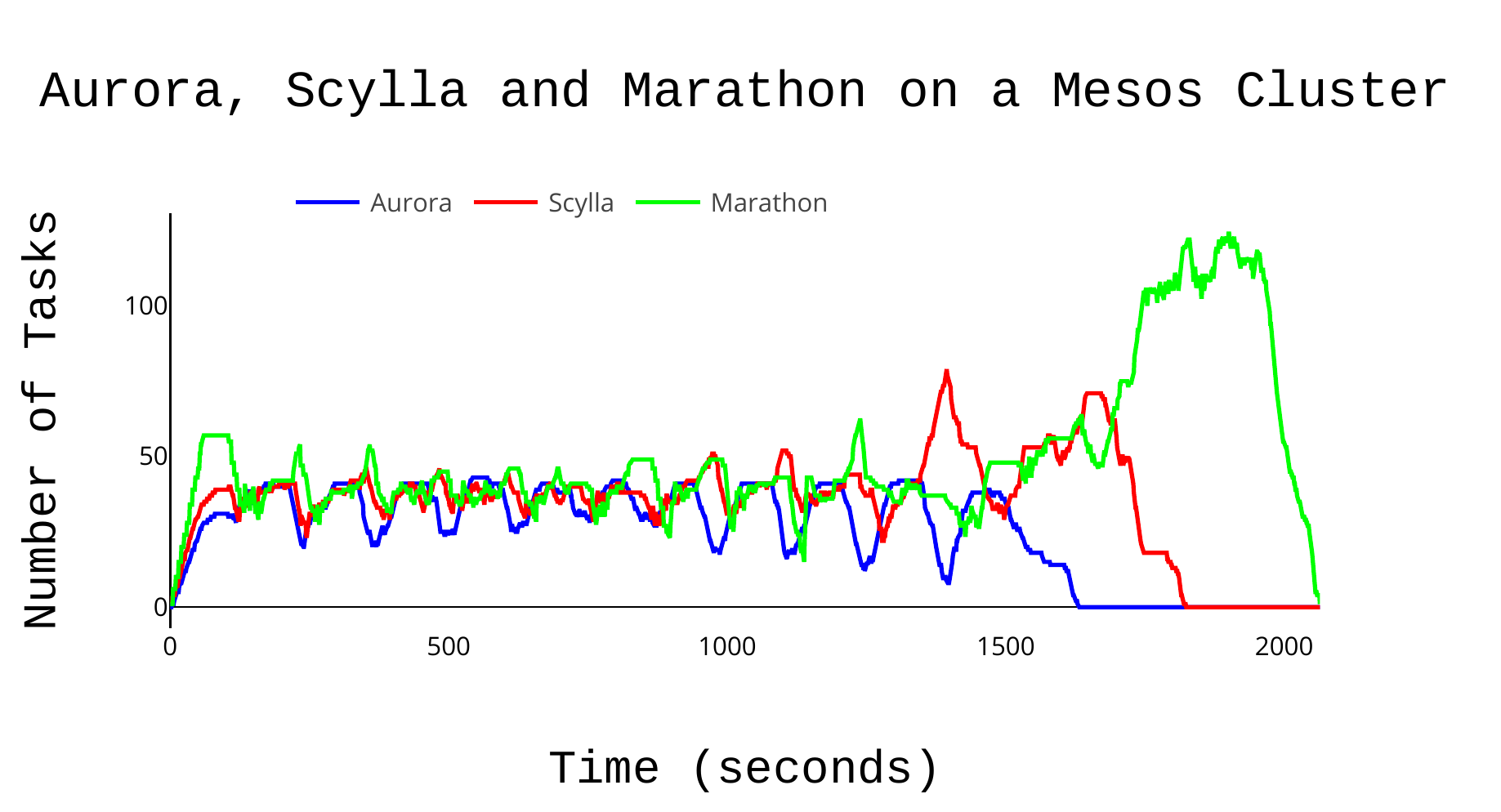}
  \caption{{\it Tromino improves the fairness by incorporating DRF-Aware policy in the task dispatcher.
  }}
  \label{fig_msa_drfaware}
   \vspace{-0.7em}
\end{figure}

Figure \ref{fig_msa_nosense} shows the fairness graphs of  Aurora, Marathon, and Scylla competing for resources. Aurora could not launch a fair number of tasks. This can be attributed both to its default configuration of holding on to offers without using them for a long period of time and the other competing framework's greedy second level scheduling. We have incorporated DRF-Aware scheduling in {\it Tromino} to address such scenarios. 

The results in Figure \ref{fig_msa_drfaware} show that each of the frameworks can launch close to a fair number of tasks, which is 42 in the cluster.
 In the following experiments, we configured {\it Tromino} with DRF awareness as the baseline to compare other {\it Tromino} policies in different case scenarios.
\subsection{\textbf{Experiment 2}: Frameworks with equal number of tasks, but with fast and slow arrival rates. }
\label{experiment2}
In this experimental setup, we have instrumented Aurora, Marathon, and Scylla to launch tasks in our experimental Mesos cluster. Tromino receives tasks for Aurora at a faster rate than Scylla, and Marathon at a slower rate than Scylla as shown in the Table~\ref{table_equalTasks}. 
All three frameworks receive an equal number of tasks to be executed. Each task is identical in terms of resource requirement as mentioned in section \ref{evaluation}. The fair number of tasks at any point in time for each framework is 42.

\begin{table}[H]
\begin{tabular}{|l|l|l|ll}
\cline{1-3}
& { \bf \# of tasks} & \begin{tabular}[c]{@{}l@{}}{ \bf Arrival Interval (sec)}\end{tabular} &  &  \\ \cline{1-3}
Aurora & 733 & 1 &  &  \\ \cline{1-3}
Marathon & 733 & 1.5 &  &  \\ \cline{1-3}
Scylla & 733 & 2 &  &  \\ \cline{1-3}
\end{tabular}
\caption{\it Configuration: Aurora, Marathon and Scylla with different task arrival rate for launching same number of tasks.}
\label{table_equalTasks}
\end{table}
Figure \ref{fig_exp2_Tromino_With_DRF_Aware}, \ref{fig_exp2_Tromino_With_Demand_Aware}, and \ref{fix_exp2_Tromino_With_DemandDRF_Aware} show the fairness plots when {\it Tromino} is configured with different task dispatching policies in the cluster. {\it Tromino} receives Aurora's tasks at a faster rate than Marathon and Scylla's tasks. During DRF-Aware policy configuration, Aurora faces a higher waiting time compared to Marathon and Scylla. Aurora is affected by a 44\% higher waiting time compared to the cluster's average for all tasks. For the Demand-Aware policy, Aurora's average waiting time is reduced by 30\% below the cluster's average. However, due to the lower task demand from Scylla, {\it Tromino} increased its waiting time to 27\% above the cluster average. For Marathon, both the policies' average waiting time stays within 10\% the cluster's average. In Demand-DRF-Aware policy, the average waiting time of the other two frameworks is within 2\% of the cluster's average. Figure \ref{fig_exp2_waiting_time_total_framework.pdf} presents the total waiting time for all three frameworks for different {\it Tromino} policies. Similarly, Figure \ref{fig_exp2_avg_waiting_time} shows and compares the average waiting time per every 100 tasks to be scheduled by each framework for each {\it Tromino} policy. Lastly, Figure \ref{fig_exp2_waiting_time_total_policy} compares the total waiting time for each policy for all the tasks in the cluster. Table\ref{table_exp2_results} provides the results for the above mentioned figures.

\begin{table}[H]
\begin{tabular}{|l|l|l|l|l}
\cline{1-4}
 & { \bf Aurora} & { \bf Marathon} & { \bf Scylla} &  \\ \cline{1-4}
DRF Aware & 44.24\% & -6.37\% & -37.87\% &  \\ \cline{1-4}
Demand Aware & -30.42\% & 2.57\% & 27.85\% &  \\ \cline{1-4}
Demand-DRF Aware & \textbf{-1.06\%} & \textbf{1.19\%} & \textbf{-0.13\%} &  \\ \cline{1-4}
\end{tabular}
\caption{{\it Result: Difference between average waiting time of each framework from average waiting time of the cluster for different Tromino policies in Experiment 2.}}
\label{table_exp2_results}
\end{table}

\begin{figure*}%
	\captionsetup[subfigure]{justification=centering}
	\centering
    \subfloat[{\it Fairness Graph when Tromino is DRF aware
    } 
    ]{{\includegraphics[width=0.30\linewidth]
    {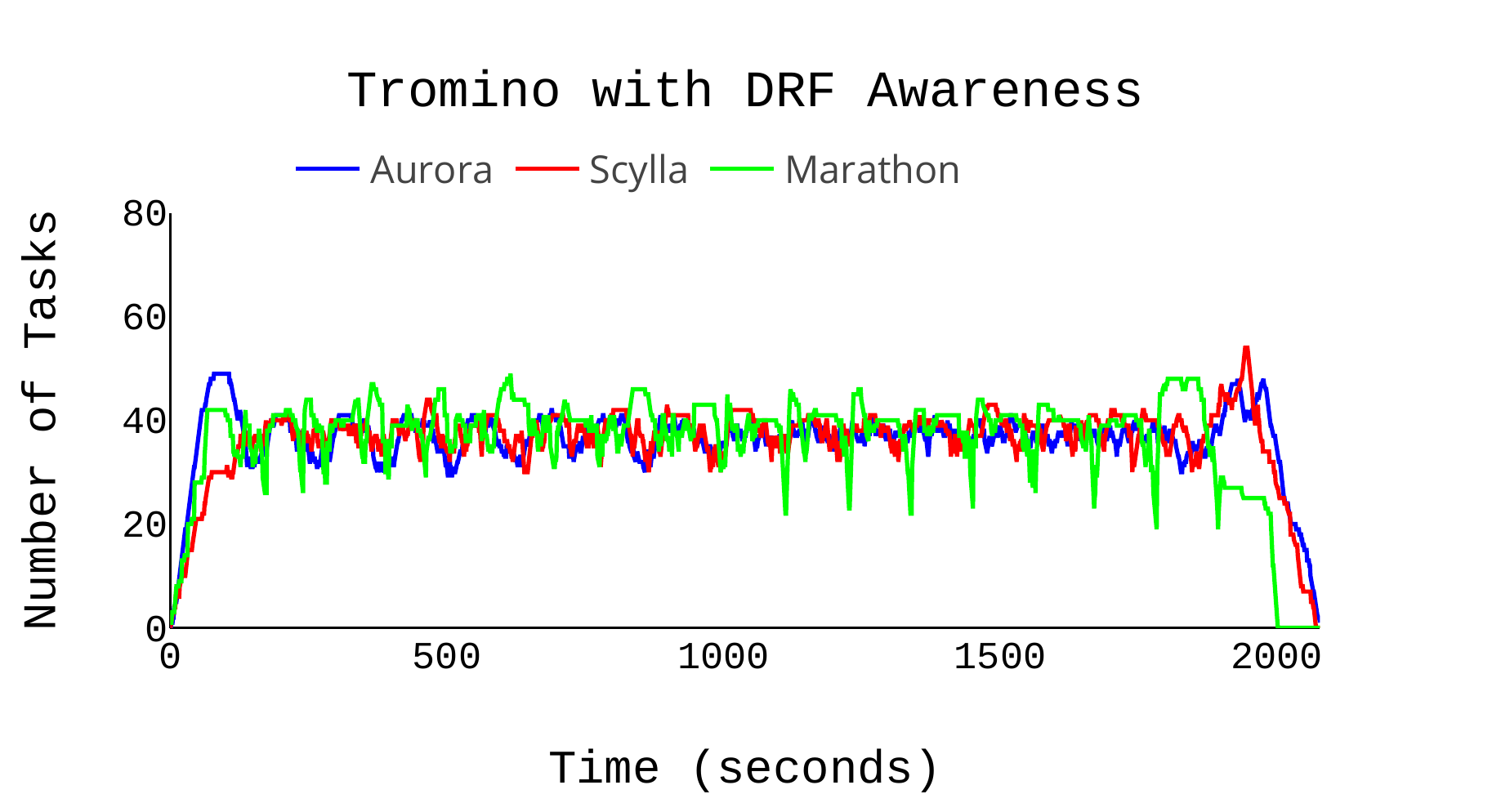}  
    \label{fig_exp2_Tromino_With_DRF_Aware}
    }}
    \subfloat[{\it  Fairness Graph when Tromino is Demand Aware
    }]
{{\includegraphics[width=0.30\linewidth]
    {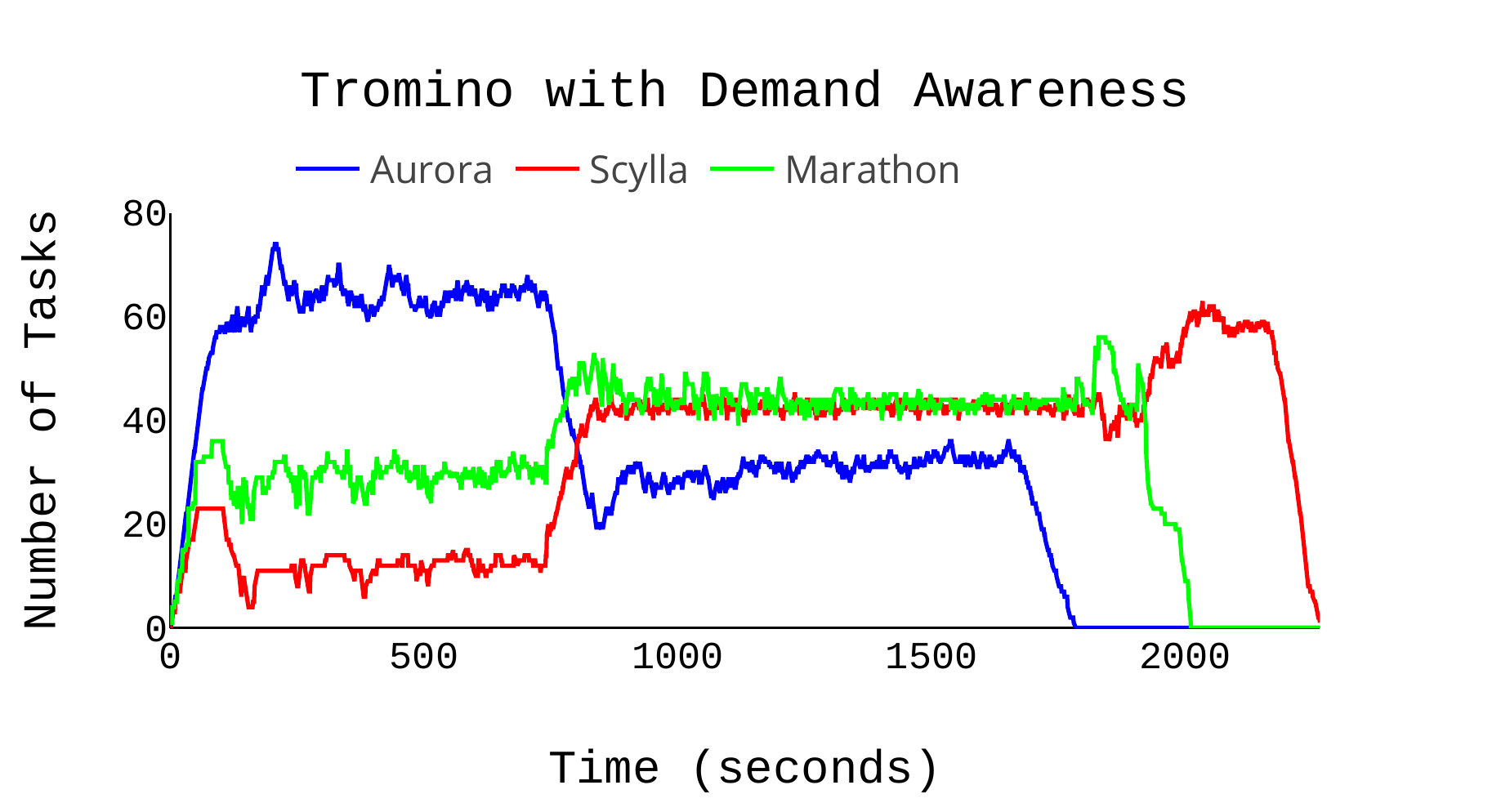}
    \label{fig_exp2_Tromino_With_Demand_Aware}
    }}
    \subfloat[{\it  Fairness Graph when Tromino is Demand and DRF aware}]
{{\includegraphics[width=0.30\linewidth]
    {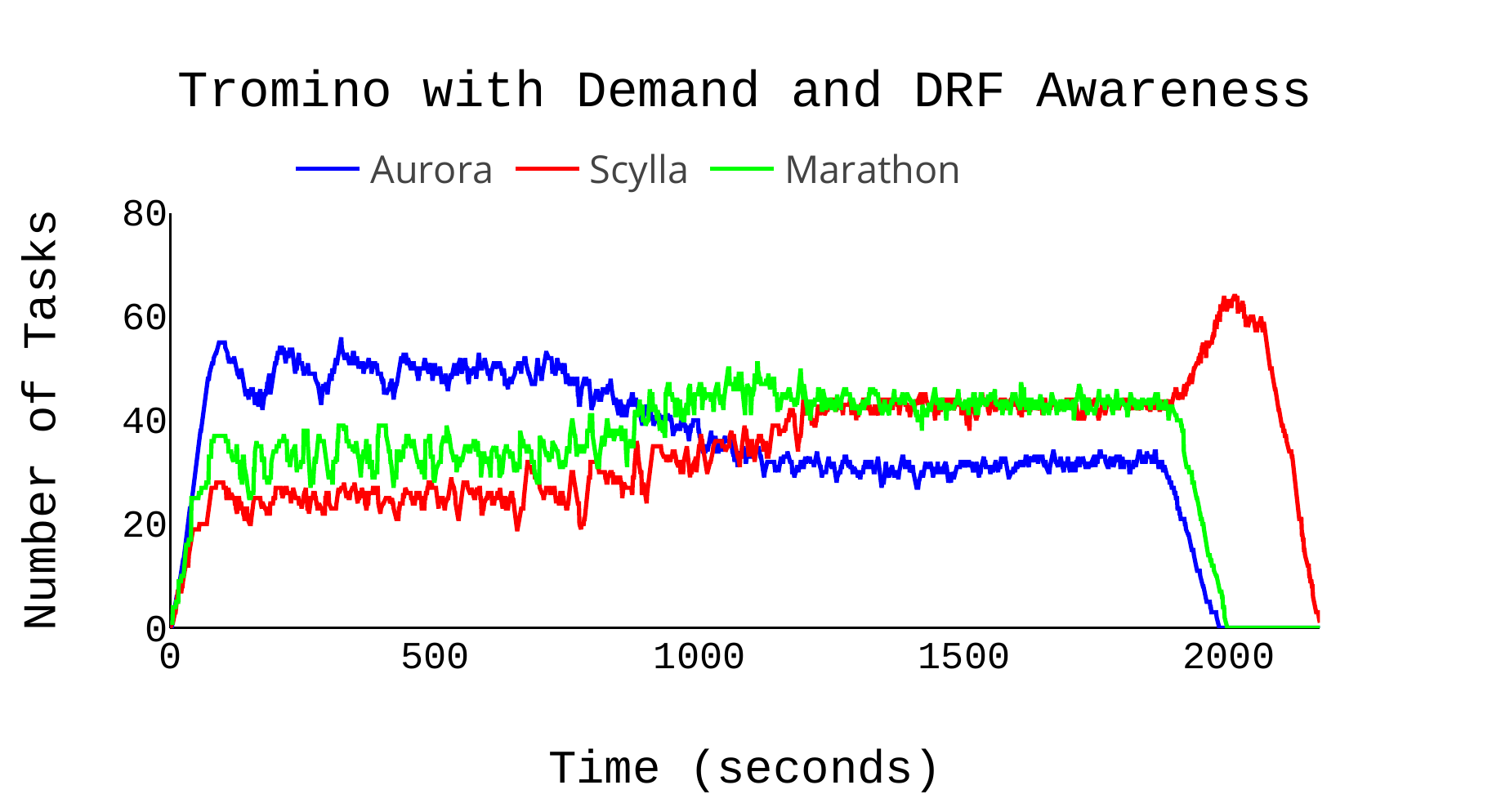}
    \label{fix_exp2_Tromino_With_DemandDRF_Aware}
    }}
\caption{{\it Resource Fairness for Experiment 2:  Results show the fairness obtained by the cluster when Tromino is configured with different policies and equal number of tasks are launched in the cluster with different task arrival rates for each framework. }}

\label{fig_equalTasks_fairnessGraphs}
\end{figure*}

\begin{figure*}%
	\captionsetup[subfigure]{justification=centering}
	\centering
    \subfloat[{\it Waiting Time for Each Framework.
    } 
    ]{{\includegraphics[width=0.30\linewidth]
    {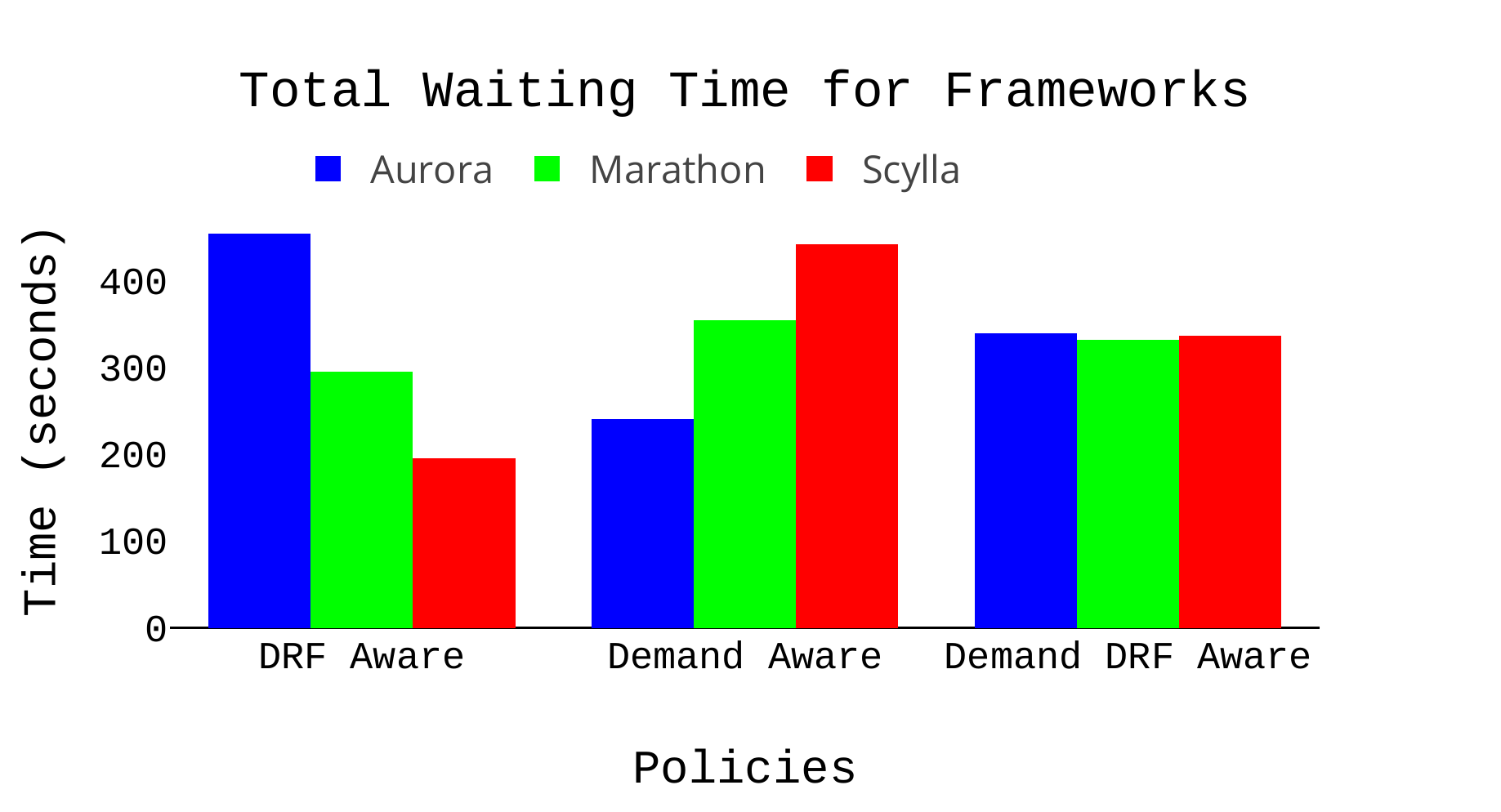}  
    \label{fig_exp2_waiting_time_total_framework.pdf}
    }}
    \subfloat[{\it  Average Waiting Time of 100 Tasks for Each Framework
    }]
{{\includegraphics[width=0.30\linewidth]
    {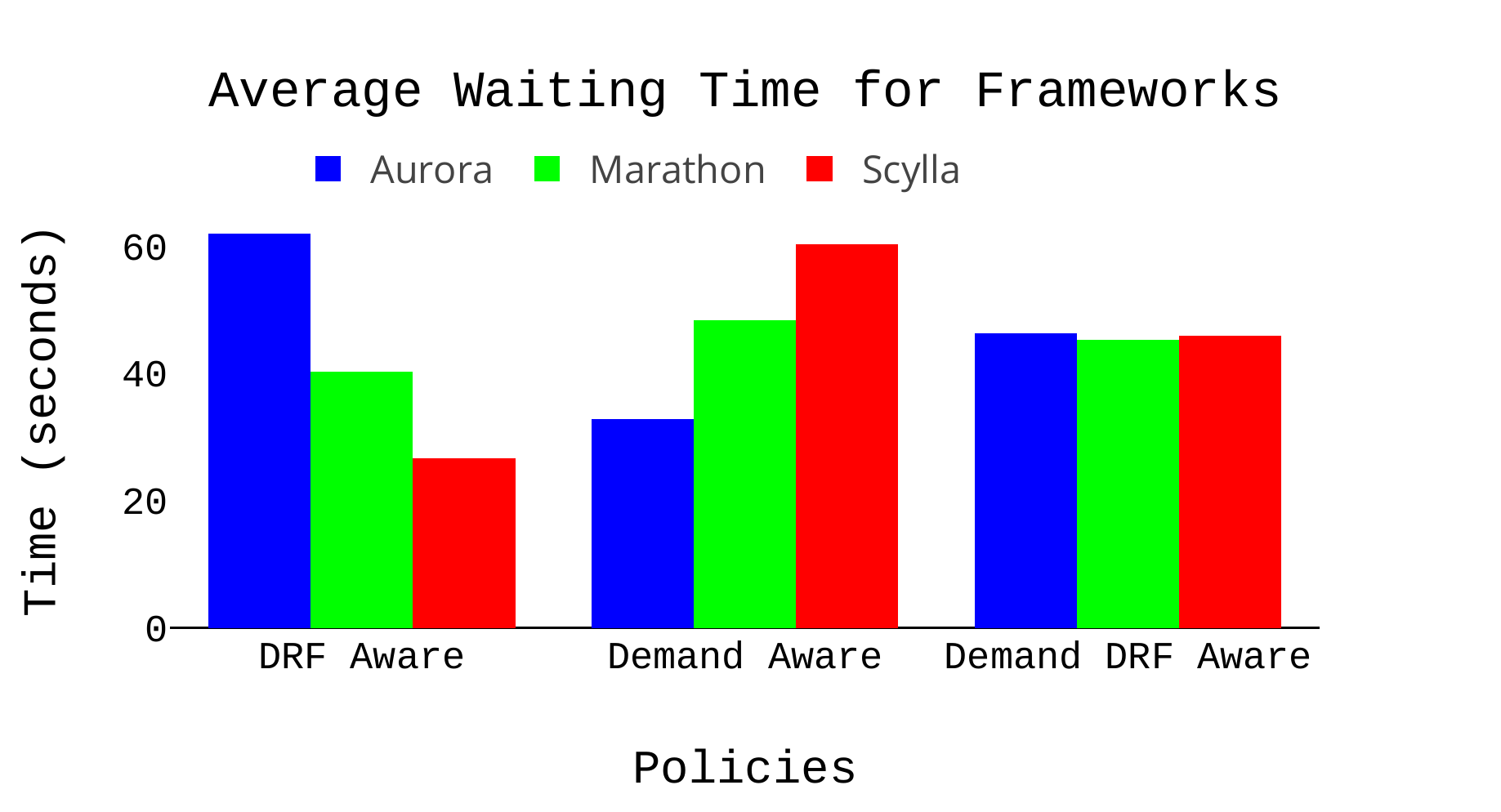}
    \label{fig_exp2_avg_waiting_time}
    }}
    \subfloat[{\it  Total Waiting Time for Each Policy}]
{{\includegraphics[width=0.30\linewidth]
    {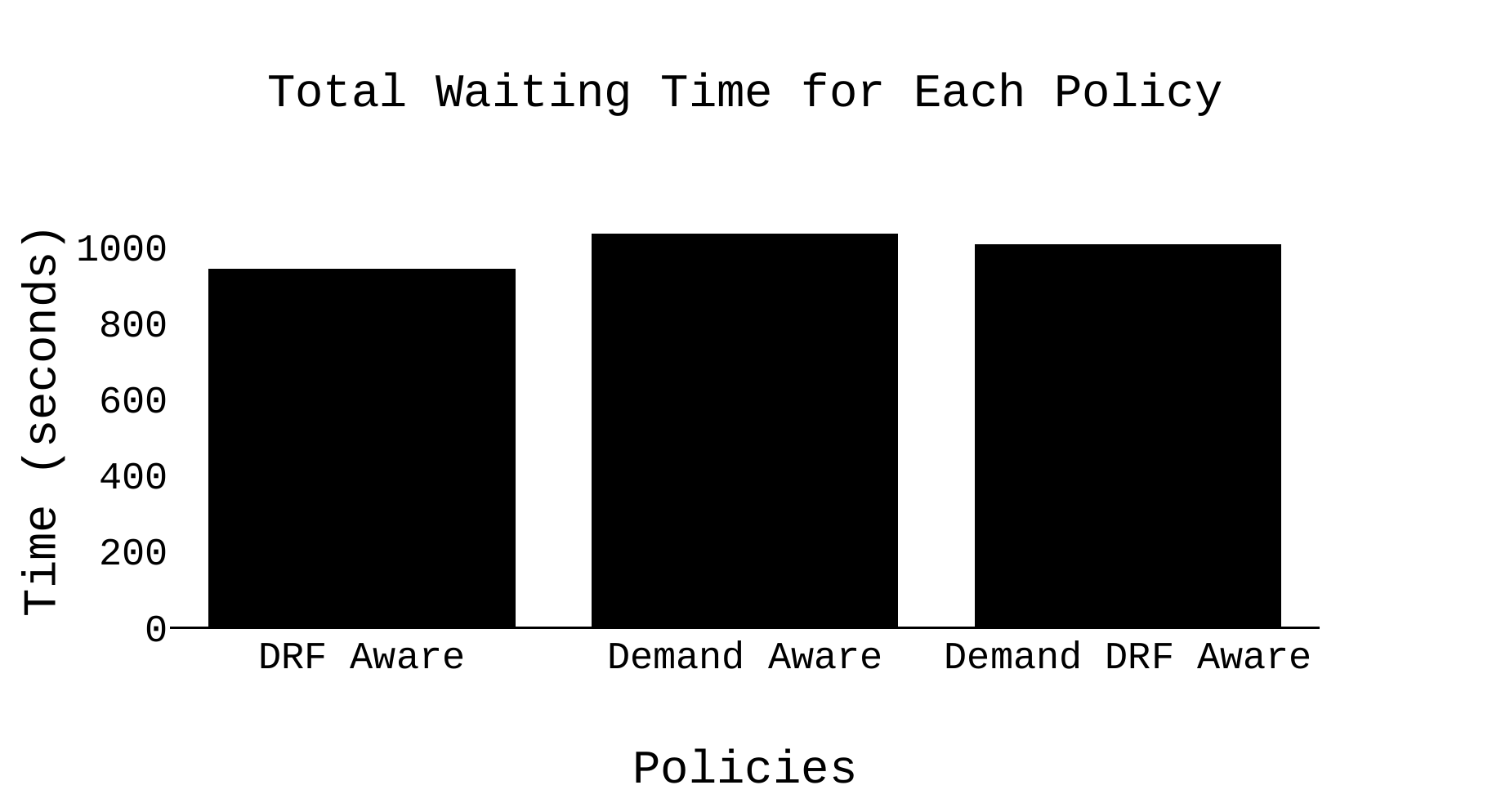}
    \label{fig_exp2_waiting_time_total_policy}
    }}
\caption{{\it Results for Experiment 2: Equal number of tasks are launched by Aurora, Marathon and Scylla in a Mesos cluster. Experimental results show how total waiting time and average waiting time varies for Aurora, Marathon and Scylla when Tromino is configured with different policies.} }

\label{fig_equalTasks_results}
\vspace{-0.5em}
\end{figure*}
\subsection{\textbf{Experiment 3}: Large number of tasks with higher arrival rates, and lower number of tasks with slower arrival rates.}
\label{experiment3}


In this experimental setup, {\it Tromino} receives fast arriving tasks for Aurora, slow arriving tasks for Scylla, and Marathon's task arrival rate is in between Aurora and Scylla's rate. Tromino receives a higher number of tasks for Aurora and fewer tasks for Scylla compared to the number of tasks received for Marathon. The task arrival rate and the number of tasks for each framework is mentioned in Table ~\ref{table_exp3}. {\it Tromino} is configured with all three policies as discussed in section \ref{policies}. In Figures \ref{fig_fairness_exp3} and \ref{fig_waiting_time_exp3}, we present our observations about resource fairness and how waiting time varies for all the policies. 

\begin{table}[H]
\begin{tabular}{|l|l|l|ll}
\cline{1-3}
& { \bf \# of tasks} & \begin{tabular}[c]{@{}l@{}} { \bf Arrival Interval (sec)} \end{tabular} &  &  \\ \cline{1-3}
Aurora & 1000 & 1 &  &  \\ \cline{1-3}
Marathon & 700 & 1.5 &  &  \\ \cline{1-3}
Scylla & 500 & 2 &  &  \\ \cline{1-3}
\end{tabular}
\caption{\it Configuration: Tromino receives more tasks and at a fast rate for Aurora, and lesser number of tasks at a slower rate for Scylla.}
\label{table_exp3}
\end{table}

Figure~\ref{fig_fairness_exp3} shows the resource fairness for all three frameworks after configuring {\it Tromino} with all three policies. In DRF aware policy configuration, Aurora's average waiting time is 73\% more than the overall average waiting time of the cluster. For Scylla, with slow arriving tasks, the waiting time is 55\% less.  After changing the configuration to follow the Demand-Aware policy, the average task waiting time difference changed to 31\% less and 34\% more for Aurora and Scylla respectively. The average waiting time difference for all three frameworks is aligned better with the cluster's average when {\it Tromino} is configured with Demand-DRF aware policy.

Figure \ref{fig_exp3_waiting_time_total_framework.pdf} presents the total waiting time for all three frameworks for different {\it Tromino} policies. Similarly, Figure \ref{fig_exp3_avg_waiting_time} shows and compares the average waiting time per every 100 tasks to be scheduled by each framework for each {\it Tromino} policy. Lastly, Figure \ref{fig_exp3_waiting_time_total_policy} compares the total waiting time for each policy for all the tasks in the cluster. Table~\ref{table_exp3_results} provides the results for the mentioned figures.

\begin{table}[H]
\begin{tabular}{|l|l|l|l|l}
\cline{1-4}
 & { \bf Aurora} & { \bf Marathon} & { \bf Scylla} &  \\ \cline{1-4}
DRF Aware & 73.33\% & -18.16\% & -55.17\% &  \\ \cline{1-4}
Demand Aware & -31.07\% & -3.30\% & 34.37\% &  \\ \cline{1-4}
Demand-DRF Aware & \textbf{2.30\%} & \textbf{-1.42\%} & \textbf{-0.88\%} &  \\ \cline{1-4}
\end{tabular}
\caption{{\it Results: Difference of average waiting time of each framework compared to average waiting time of the cluster for different Tromino policies in Experiment 3.}}
\label{table_exp3_results}
\end{table}

\begin{figure*}%
	\captionsetup[subfigure]{justification=centering}
	\centering
    \subfloat[{\it DRF Aware.
    } 
    ]{{\includegraphics[width=0.30\linewidth]
    {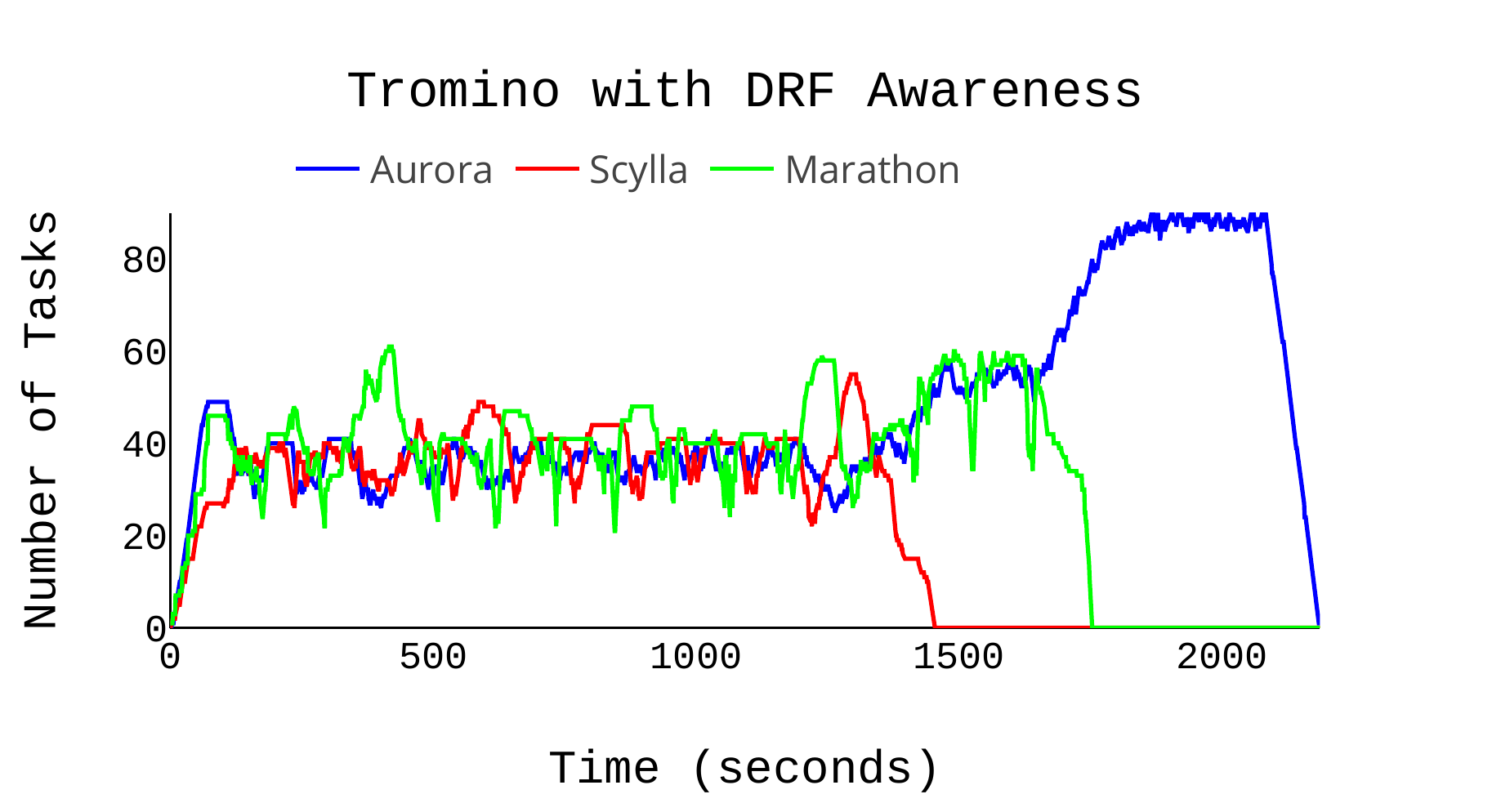}  
    \label{fig_exp3_Tromino_With_DRF_Aware}
    }}
    \subfloat[{\it  Demand Aware
    }]
{{\includegraphics[width=0.30\linewidth]
    {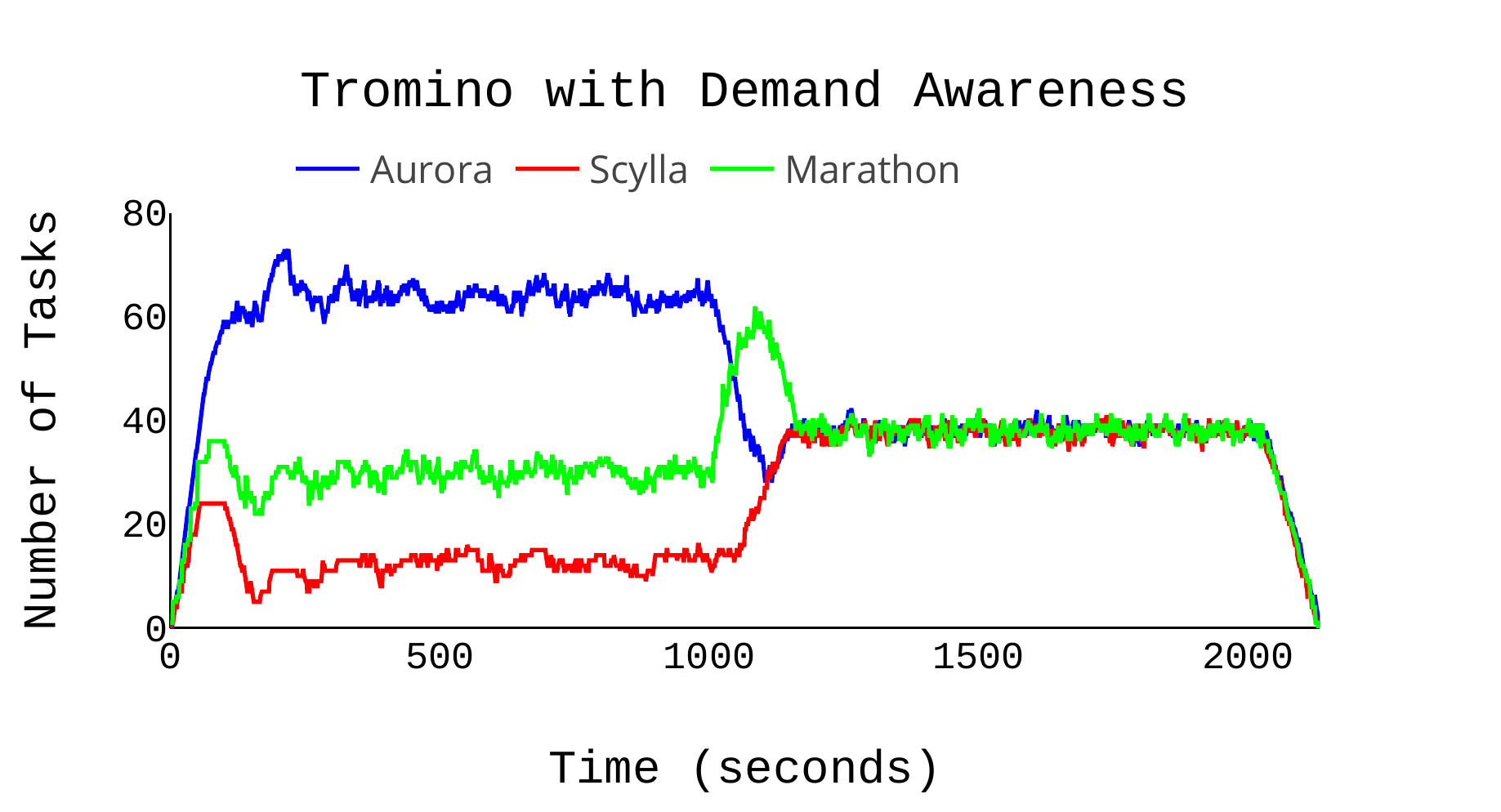}
    \label{fig_exp3_Tromino_With_Demand_Aware}
    }}
    \subfloat[{\it  Demand and DRF aware.}]
{{\includegraphics[width=0.30\linewidth]
    {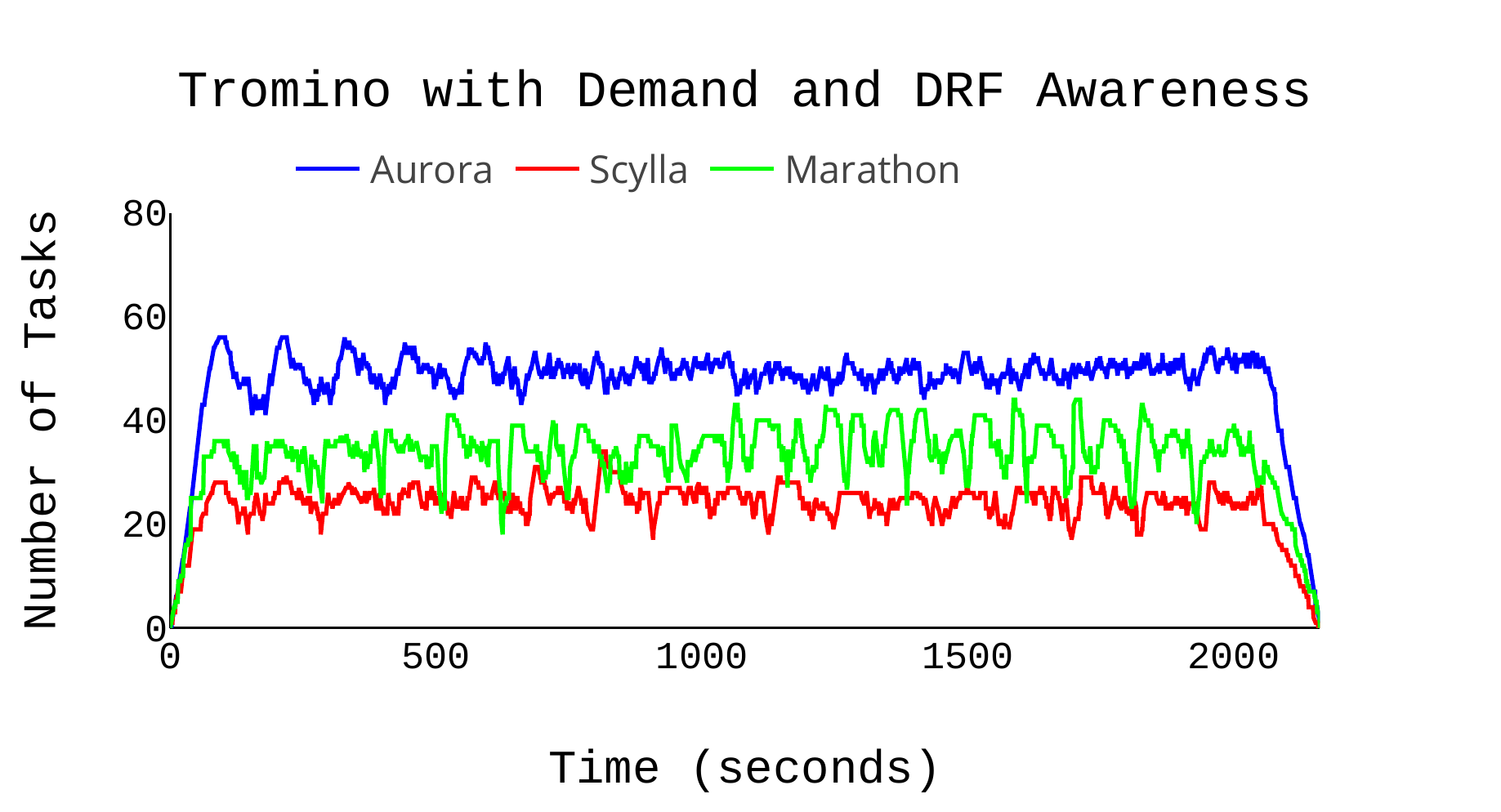}
    \label{fig_exp3_Tromino_With_DemandDRF_Aware}
    }}
\caption{{\it Resource Fairness for Experiment 3: Resource fairness obtained when Tromino is configured for different policies. Tromino receives more tasks for Aurora at a fast rate whereas Scylla's tasks arrive at a slower rate and are lesser in number.}}

\label{fig_fairness_exp3}
\end{figure*}


\begin{figure*}%
	\captionsetup[subfigure]{justification=centering}
	\centering
    \subfloat[{\it Waiting Time for Each Framework.
    } 
    ]{{\includegraphics[width=0.30\linewidth]
    {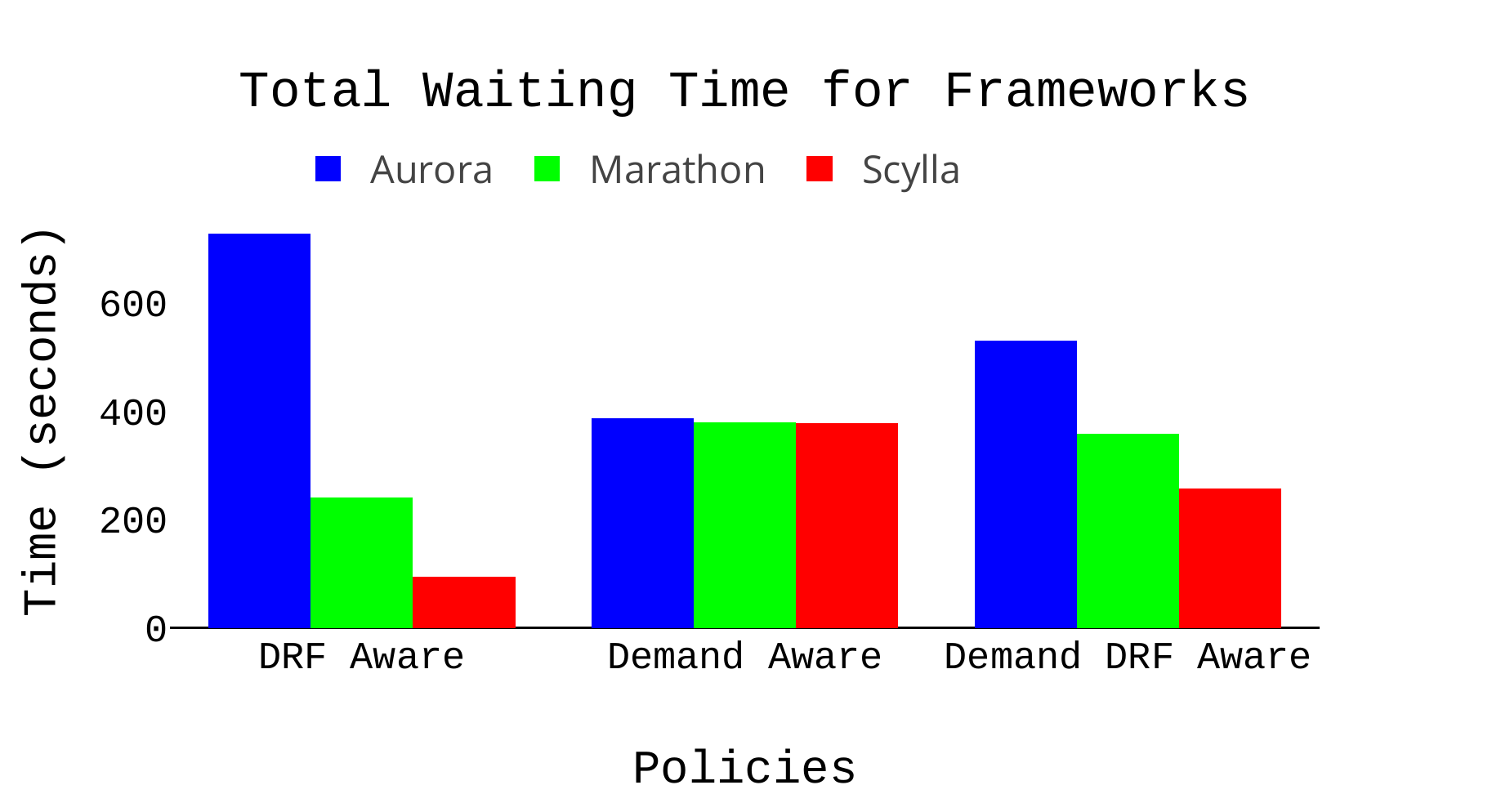}  
    \label{fig_exp3_waiting_time_total_framework.pdf}
    }}
    \subfloat[{\it  Average Waiting Time for Each Framework
    }]
{{\includegraphics[width=0.30\linewidth]
    {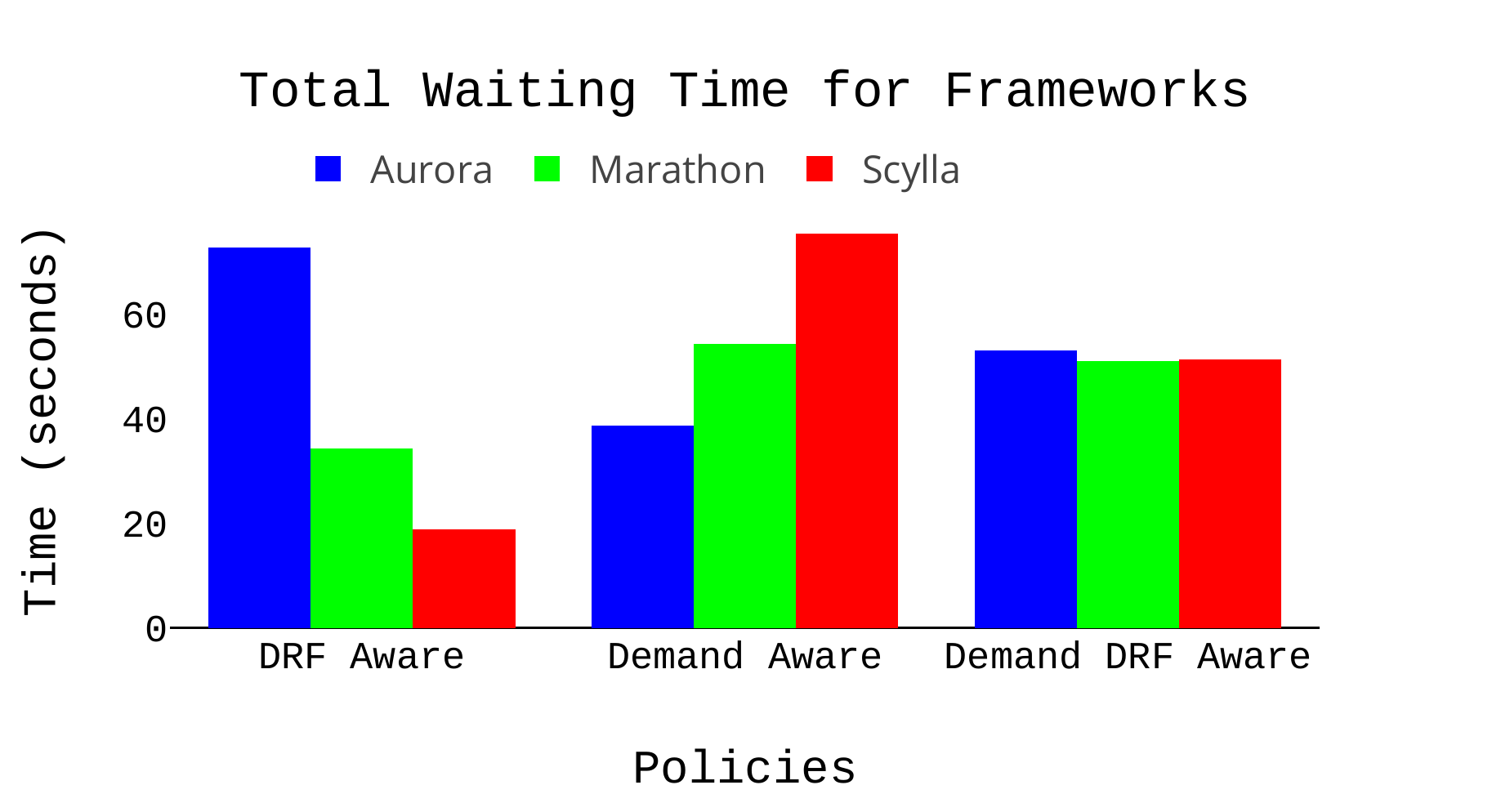}
    \label{fig_exp3_avg_waiting_time}
    }}
    \subfloat[{\it  Total Waiting Time of 100 Tasks for Each Policy}]
{{\includegraphics[width=0.30\linewidth]
    {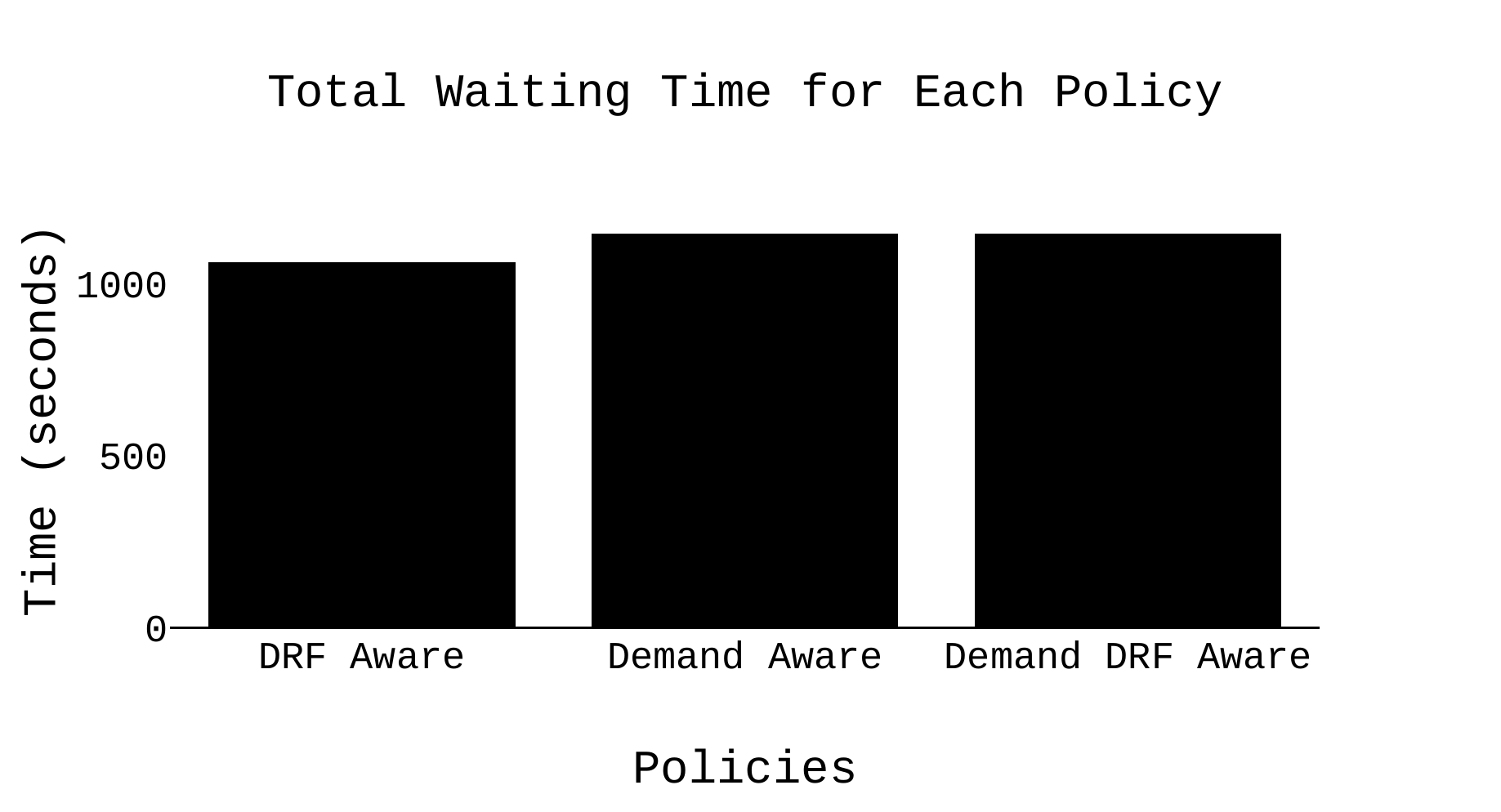}
    \label{fig_exp3_waiting_time_total_policy}
    }}
\caption{{\it Results for Experiment 3:  Results show how total waiting time and average waiting time varies for Aurora, Marathon and Scylla when Tromino is configured with different policies. Tromino receives higher number of tasks for Aurora in a higher arrival rate than Marathon and Scylla (Table: \ref{table_exp3}).}}
\label{fig_waiting_time_exp3}
\end{figure*}
\subsection{\textbf{Experiment 4}: Large number of tasks with slower arrival rates, and lower number of tasks with faster arrival rates.}

In this experimental setup, a fewer number of Aurora tasks are received by {\it Tromino} at a faster arrival rate, and unlike the previous experimental setup, {\it Tromino} receives more Scylla tasks at a slower rate. In Table \ref{table_exp4}, we present the number of tasks received for each framework and the arrival rate. 

\begin{table}[H]
\begin{tabular}{|l|l|l|ll}
\cline{1-3}
& { \bf \# of tasks} & \begin{tabular}[c]{@{}l@{}}{ \bf Arrival Interval (sec)}\end{tabular} &  &  \\ \cline{1-3}
Aurora & 500 & 1 &  &  \\ \cline{1-3}
Marathon & 700 & 1.5 &  &  \\ \cline{1-3}
Scylla & 900 & 2 &  &  \\ \cline{1-3}
\end{tabular}
\caption{\it Configuration: Fewer tasks, but at a faster rate, for Auroras tasks; and a larger number of slow arriving Scylla tasks for Experiment 4.}
\label{table_exp4}
\end{table}
Figure \ref{fig_exp4_waiting_time_total_framework.pdf} presents the total waiting time for all three frameworks for different Tromino policies. Similarly, Figure \ref{fig_exp4_avg_waiting_time} shows and compares the average waiting time per every 100 tasks to be scheduled by each framework for each {\it Tromino} policy. Lastly, Figure \ref{fig_exp4_waiting_time_total_policy} compares the total waiting time for each policy for all the tasks in the cluster. Table \ref{table_exp4_results} shows the difference of average waiting time of each framework with each policy configuration compared to the overall cluster's average waiting time.

\begin{table}[H]
\begin{tabular}{|l|l|l|l|l}
\cline{1-4}
& { \bf Aurora} & { \bf Marathon} & { \bf Scylla} &  \\ \cline{1-4}
DRF Aware & 16.67\% & 7.61\% & -24.28\% &  \\ \cline{1-4}
Demand Aware & -35.93\% & 8.78\% & 27.15\% &  \\ \cline{1-4}
Demand-DRF Aware & \textbf{-10.70\%} & \textbf{4.03\%} & \textbf{6.67\%} &  \\ \cline{1-4}
\end{tabular}
\caption{{}\it Result: Difference of average waiting time of each framework compared to average waiting time of the cluster for different Tromino policies in Experiment 4.}
\label{table_exp4_results}
\end{table}

\begin{figure*}%
	\captionsetup[subfigure]{justification=centering}
	\centering
    \subfloat[{\it DRF Aware.
    } 
    ]{{\includegraphics[width=0.30\linewidth]
    {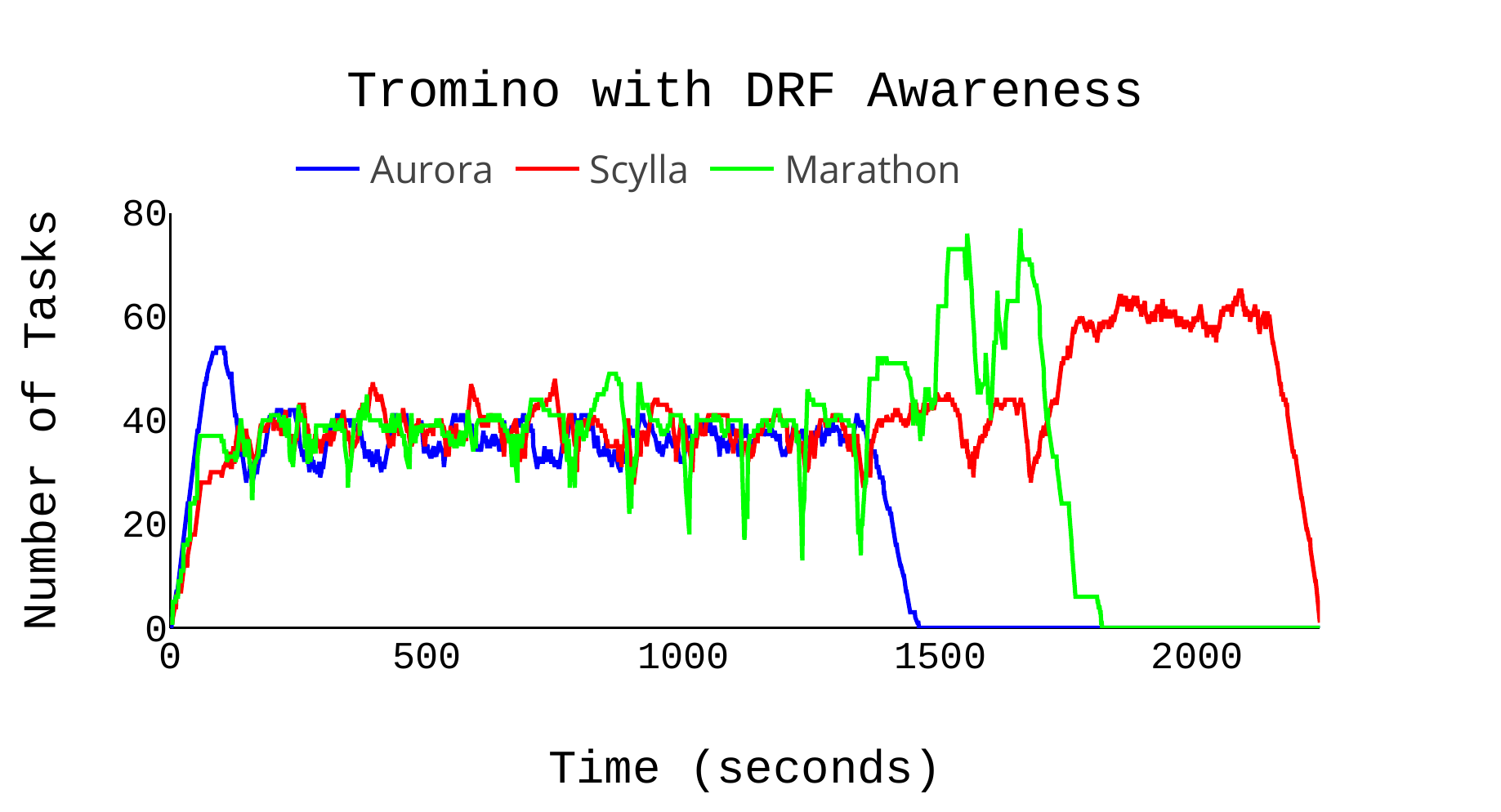}  
    \label{fig_exp4_Tromino_With_DRF_Aware}
    }}
    \subfloat[{\it  Demand Aware
    }]
{{\includegraphics[width=0.30\linewidth]
    {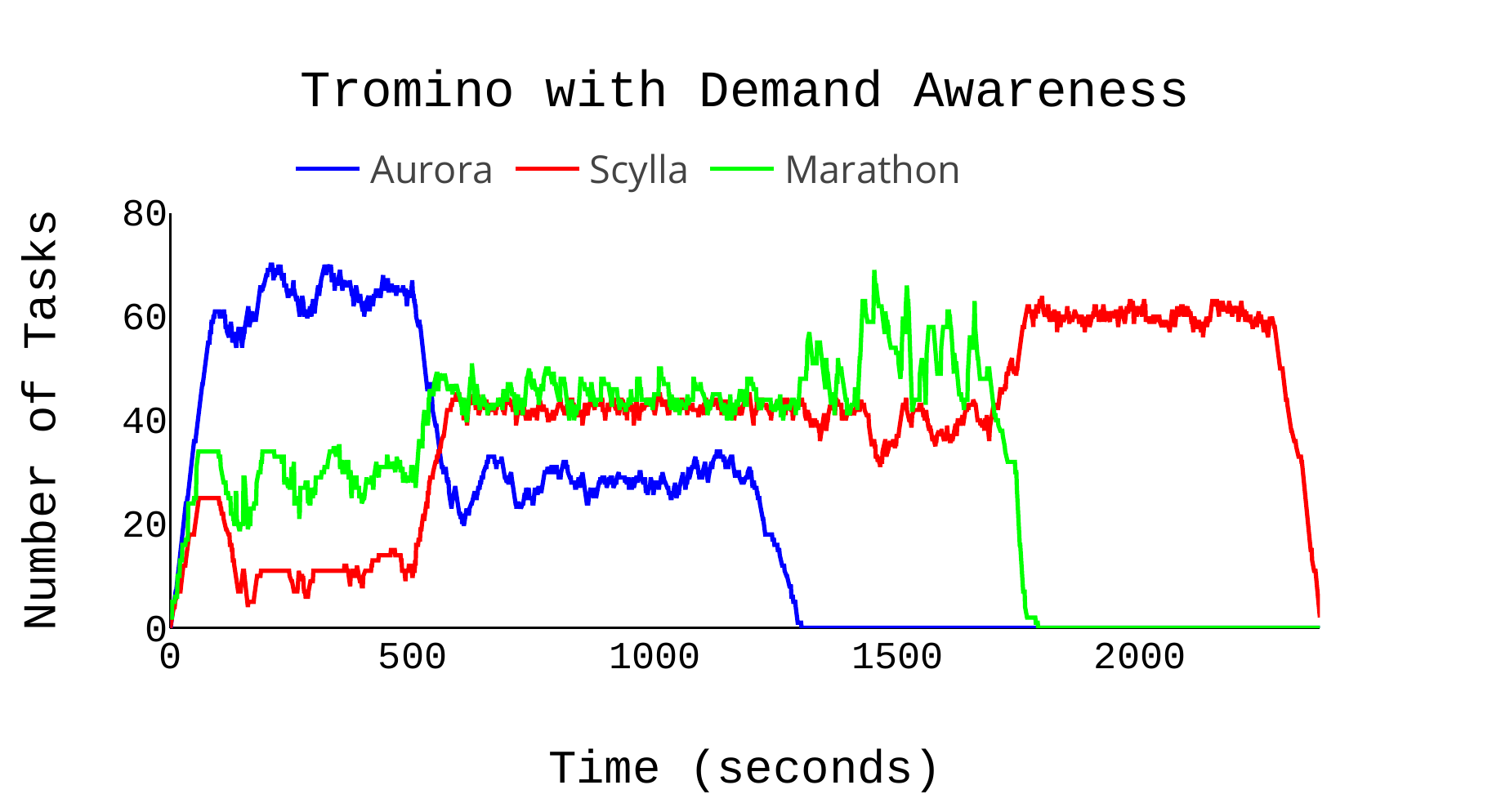}
    \label{fig_exp4_Tromino_With_Demand_Aware}
    }}
    \subfloat[{\it  Demand and DRF aware.}]
{{\includegraphics[width=0.30\linewidth]
    {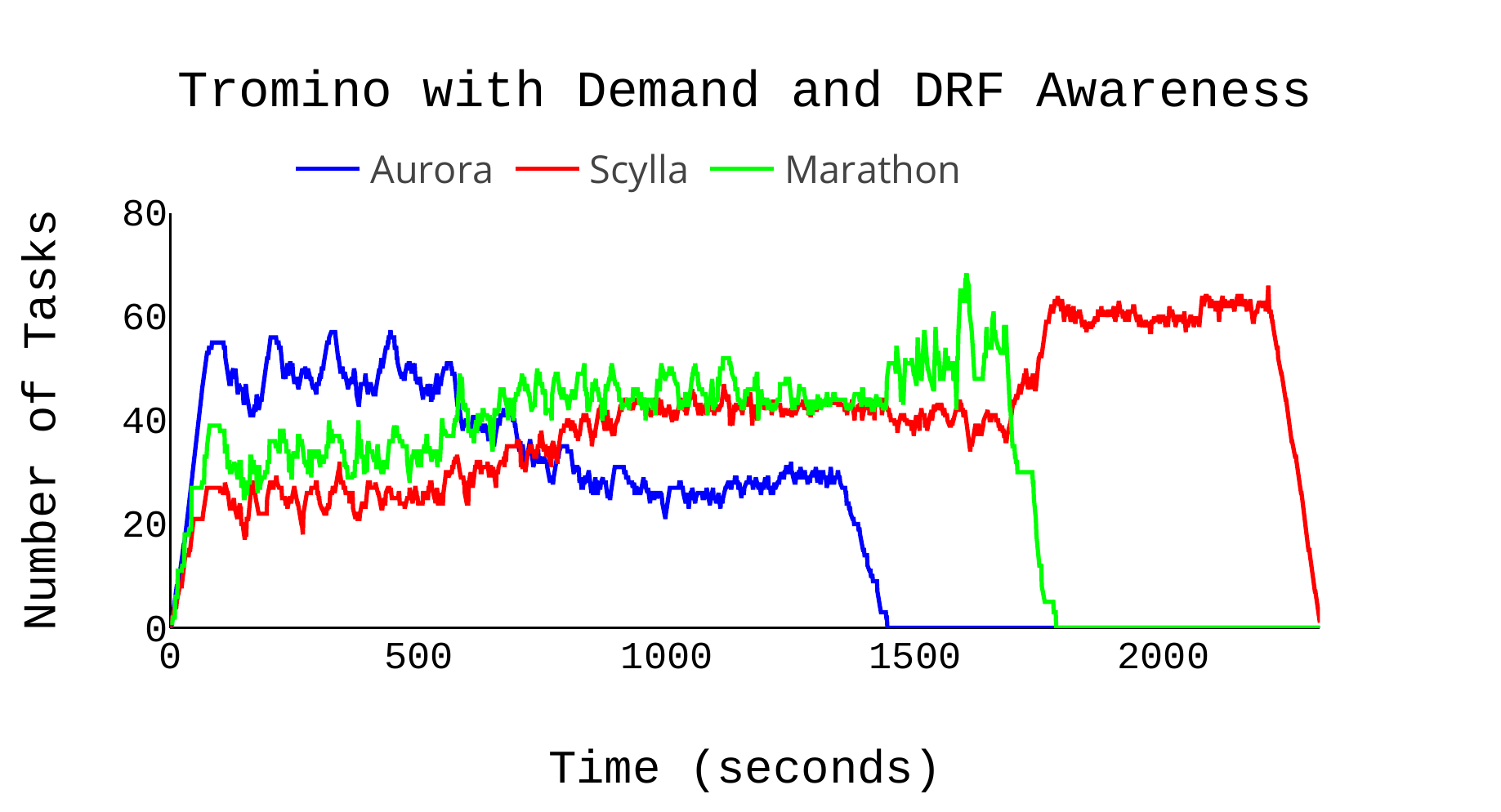}
    \label{fig_exp4_Tromino_With_DemandDRF_Aware}
    }}
\caption{{\it Resource Fairness for Experiment 4: Resource fairness of the Mesos cluster for different Tromino policies. Tromino receives fewer tasks for Aurora at a fast rate whereas Scylla's tasks arrive at a slower rate but are more in number.}}
\end{figure*}

\vspace{-0.7em}
\begin{figure*}%
	\captionsetup[subfigure]{justification=centering}
	\centering
    \subfloat[{\it Waiting Time for Each Framework.
    } 
    ]{{\includegraphics[width=0.30\linewidth]
    {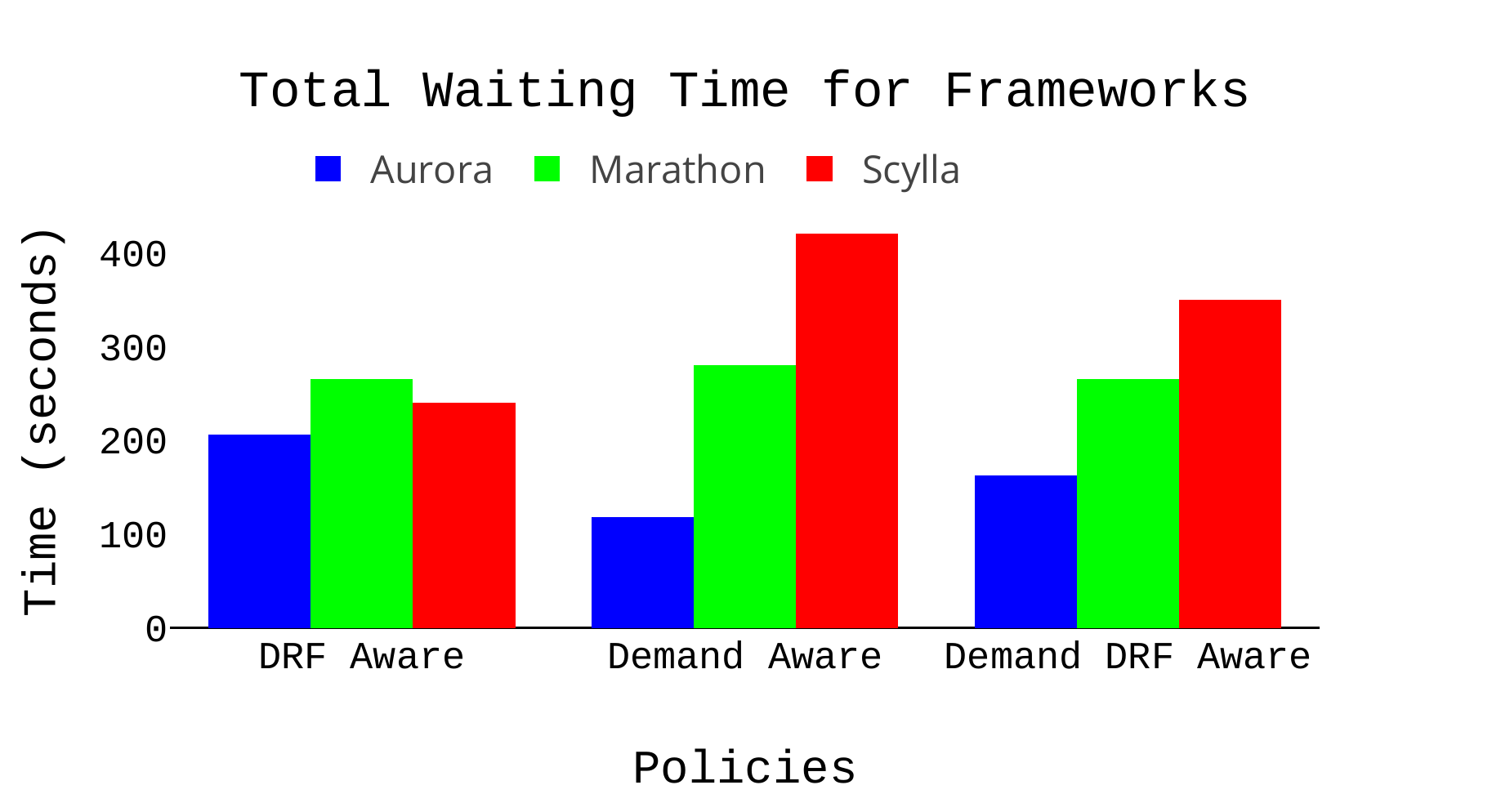}  
    \label{fig_exp4_waiting_time_total_framework.pdf}
    }}
    \subfloat[{\it  Average Waiting Time of 100 Tasks for Each Framework
    }]
{{\includegraphics[width=0.30\linewidth]
    {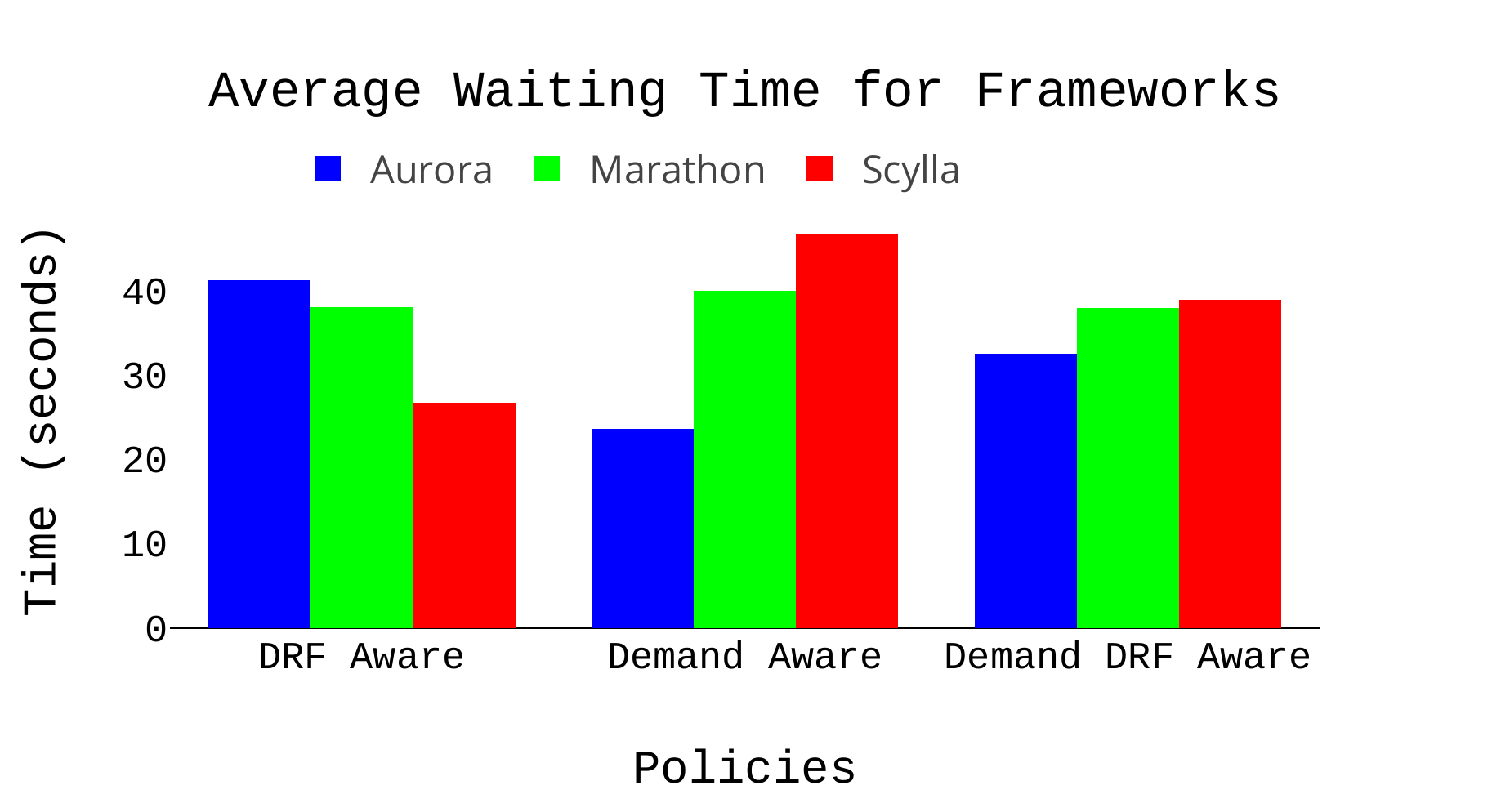}
    \label{fig_exp4_avg_waiting_time}
    }}
    \subfloat[{\it  Total Waiting Time for Each Policy}]
{{\includegraphics[width=0.30\linewidth]
    {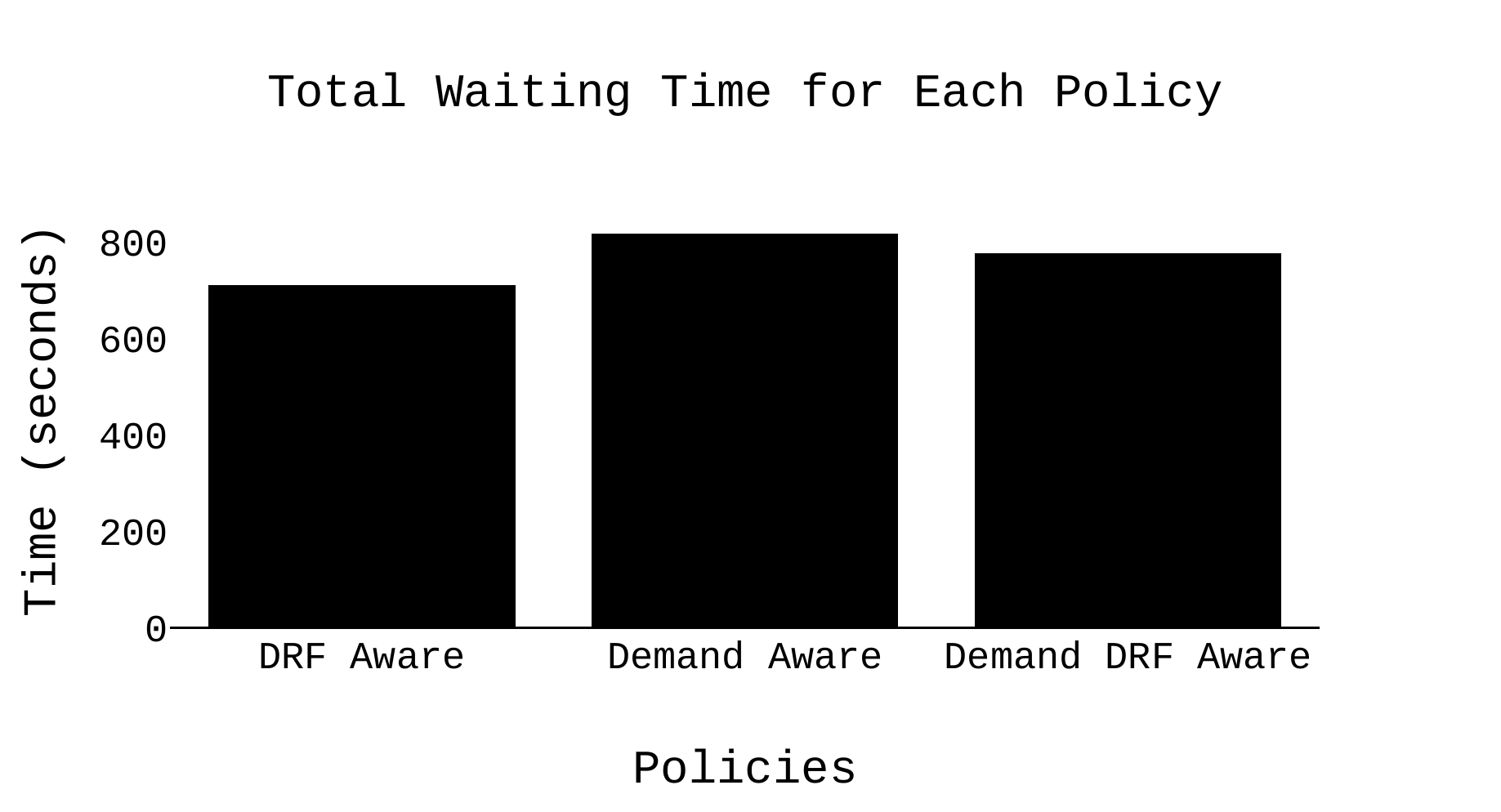}
    \label{fig_exp4_waiting_time_total_policy}
    }}
\caption{ {\it Results for Experiment 4:  Results show how total waiting time and average waiting time varies for Aurora, Marathon and Scylla when Tromino is configured with different policies and tasks arrive at different arrival rates (Table~\ref{table_exp4}). }}
\end{figure*}

\section{related work}
We have proposed a few policies to evaluate the fairness of an Apache Mesos cluster based on average waiting time. 
In our previous work \cite{saha_airavata_mesos_paper} \cite{saha_nat_mesos_paper} we have shown how Apache Mesos can be integrated with scientific workflow managers like Apache Airavata~\cite{the_airavata_paper} to run science application through Docker containers \cite{saha_docker_evaluation_paper}. The community can take advantage of Mesos based fairness to distribute resources across users. 

Khaled et al.~\cite{almi2017resource} designed and developed Resource Demand Aware Scheduling (RDAS) for scientific workflows to reduce the overall completion time. RDAS considers the structure of workflows and based on the resource demands of each stage it tries to optimize the resource allocation for better throughput. However, in our Mesos cluster, we have considered short living tasks from different users with specific resource requirements and scheduled them based on the overall demand from each user.  

Boyang et al.~\cite{peng2015r} developed {\it {R-Storm}}, which is aware of the resource demand and availability in a Storm based stream processing environment to increase the overall throughput of the cluster. Multiple Storm applications in a cluster yield better performance in the presence of R-Storm than the default Apache Storm configuration.
Fahad R et al.~\cite{dogar2014decentralized} developed {\it {Baarat}}, a task aware scheduler over the network, which dynamically schedules multiple tasks together based on the task's network bandwidth requirements. It dynamically changes the level of multiplexing in the network to optimize the average and tail completion time for data center applications.

\section{conclusion}
\begin{itemize}
\item Individual framework configuration and attributes such as offer holding period and second level scheduling policy can impose unfairness in a Mesos cluster. DRF aware task dispatching by {\it Tromino} can overcome the unfairness and establish better fairness distribution in the cluster. 

\item A Framework with a higher task demand needs to get more resources than a framework with a lesser demand to keep the overall waiting time low. {\it Tromino} can schedule tasks based on the resource demand and current resource consumption of frameworks in the cluster.

\item We orchestrated frameworks with different task arrival rates and different number of tasks to execute. Demand awareness is vital to optimize the average waiting time for each framework. 

\item Demand and DRF awareness on top of Mesos' default DRF based resource allocation can decrease the average waiting time for a framework.
\end{itemize}

\section{future work}
In the scope of our current work, we have developed a queue manager, {\it Tromino}, external to Apache Mesos, which can dispatch tasks based on the dominant share and demands by monitoring the cluster information and pending task queues. Mesos' allocation module does consider the resource demands from each user. However, it can be extended and a new allocation module can be designed that checks the available resources on each agent and can allocate resources based on the demands. In a production environment, where thousands of nodes are configured, scanning through all the nodes with its available resources can take longer time for a single allocation cycle. It will be useful to study the trade-offs between demand aware allocation for meeting better resource constraints against current random resource allocation that may not fit and wait for future cycles to get a better allocation. 

We would like to investigate and develop new policies to consider not only total resource demands but also the demand of each task at a more finer granularity. Tasks of different frameworks can have different resource requirements that can differ in the magnitude and may come with priorities. We would like to develop and test new policies to consider all such task constraints of a data center for further improvement and better resource allocation.
\section{Acknowledgements}
This work is partially supported by National Science Foundation, through the OAC-1740263 award.

\bibliographystyle{IEEEtran}
\bibliography{bib.bib}

\end{document}